\documentclass[
12pt,
3p,
review,
]{elsarticle}
\usepackage{graphicx,graphics}
\usepackage{placeins}
\usepackage{amsmath}
\usepackage{amssymb}
\usepackage{mathtools}
\usepackage{array}
\usepackage{textgreek}
\usepackage{color}
\usepackage{soul}
\usepackage{dcolumn}% Align table columns on decimal point
\usepackage{bm}% bold math
\usepackage{todonotes}
\usepackage{onimage}
\usepackage[utf8]{inputenc}
\usepackage[english]{babel}
\usepackage{libertine}
\usepackage[caption=false]{subfig}
\usepackage{pgfplots}
\usepgfplotslibrary{polar}
\pgfplotsset{compat=1.9}
\usetikzlibrary{calc}

\usepackage{multirow}

\usepackage{layouts}

\makeatletter
\def\@seccntformat#1{\@ifundefined{#1@cntformat}%
   {\csname the#1\endcsname\quad}%      default
   {\csname #1@cntformat\endcsname}%    enable individual control
}
\makeatother

\newcommand{\vphi}{\mbox{{\bm{$\phi$}}}}

\newcommand{\vc}{\mbox{{\bm{$c$}}}}

\newcommand{\vv}[1]{\boldsymbol{#1}}
\newcommand{\n}{\vv{\nabla}}
\renewcommand{\l}{\mathopen{}\mathclose\bgroup\left}
\renewcommand{\r}{\aftergroup\egroup\right}

\newcommand{\diff} [1]{\mathrm{d}{#1}} 
\newcommand{\phia}{\phi_{\alpha}}
\newcommand{\phib}{\phi_{\beta}}
\newcommand{\gab}{\gamma_{\alpha \beta}}

\newcommand{\qab}{q_{\alpha \beta}}

\let\originaleps=\epsilon
\let\epsilon=\varepsilon
\let\varepsilon=\originaleps
\usepackage{stmaryrd}

\definecolor{mydark_blue}{RGB}{0, 0, 139}
\definecolor{myblue}{RGB}{0, 0, 255}
\definecolor{mycyan}{RGB}{0, 255, 255}  
\definecolor{mygreen}{RGB}{0, 255, 0}
\definecolor{myyellow}{RGB}{255, 255, 0}
\definecolor{myred}{RGB}{255, 0, 0}
\definecolor{mydark_red}{RGB}{139, 0, 0}
\definecolor{myblack}{RGB}{0, 0, 0}

\definecolor{BRY_1}{RGB}{  0,  0,255}
\definecolor{BRY_2}{RGB}{127,  0,127}
\definecolor{BRY_3}{RGB}{255,  0,  0}
\definecolor{BRY_4}{RGB}{255,127,  0}
\definecolor{BRY_5}{RGB}{255,255, 85}

\newcommand{\calF}{\mathcal{F}} % Free energy functional

\journal{Journal of Material Science and Technology}

\begin{document}

\begin{frontmatter}
\title{Morphological stability of three-dimensional cementite rods in polycrystalline system: A phase-field analysis}

\author[mymainaddress,mysecondaryaddress]{Tobias Mittnacht}
\author[mymainaddress]{P G Kubendran Amos\corref{mycorrespondingauthor}}
\cortext[mycorrespondingauthor]{Prince Gideon Kubendran Amos}
\ead{prince.amos@kit.edu}
\author[mymainaddress,mysecondaryaddress]{Daniel Schneider}
\author[mymainaddress,mysecondaryaddress]{Britta Nestler}

% \fntext[fn1]{The authors contributed equally.}

\address[mymainaddress]{Institute of Applied Materials (IAM-CMS), Karlsruhe Institute of Technology (KIT),\\
Strasse am Forum 7, 76131 Karlsruhe, Germany
}
\address[mysecondaryaddress]{Institute of Digital Materials Science (IDM), Karlsruhe University of Applied Sciences,\\
Moltkestr. 30, 76133 Karlsruhe, Germany}

\begin{abstract} 

Transformations accompanying shape-instability govern the morphological configuration and distribution of the phases in a microstructure.
Owing to the influence of the microstructure on the properties of a material, in the present work, the stability of three-dimensional rods in a \lq representative\rq \thinspace polycrystalline system is extensively analysed.
A multiphase-field model, which recovers the physical laws and sharp-interface relations, and includes grain boundary diffusion, is adopted to investigate the morphological evolution of the precipitate.
Moreover, the efficiency of the numerical approach is ensured by establishing the volume-preserving chemical equilibrium through the incorporation TCFe8 (CALPHAD) data and solving phase-field evolution in the Allen-Cahn framework.
The morphological evolution of the rod in the representative multiphase system exhibits a unique transformation mechanism which is significantly different from the evolution of an isolated finite-structure.
It is realised that, in a polycrystalline arrangement, irrespective of the initial size of the rod, the shape-change begins with the energy-minimising events at the triple junctions.
This early transformation renders a characteristic morphology at the longitudinal ends of the structure, which introduces sufficient driving-force through the curvature-difference for the subsequent morphological changes.
The continued mass transfer to the terminations, ultimately, breaks-off the rod into separate entities that are entangled in the grain boundary.
With increase in the aspect ratio of the rod, it is identified that the source of mass transfer, which turns into the ovulation site, shifts from the centre.
This increases the number of fragmentation events and introduces satellite particle.
The size of the satellite particle is dictated by a definite ovulation criterion, which is ascertained by examining the transformation of different-sized rods.
A comprehensive understanding of the transformation kinetics and mechanism governing the morphological evolution of the rods in a polycrystalline system is rendered in this work.

\end{abstract}


\end{frontmatter}

% \linenumbers

\section{Introduction}

Critical properties like crack resistance are governed by the morphological configuration of the phases in a microstructure, in addition to their volume fractions, chemical composition and crystal structure~\cite{becher1991microstructural}.
For instance, when compared to the spheroidal distribution of precipitate in a matrix, the crack propagation path is noticeably different in a microstructure with seemingly continuous arrangement of the phases like lamellar structure~\cite{rao1995fatigue,tokaji2007roles}.
This disparity can be observed despite the similarity in the phase fractions.
Therefore, a comprehensive understanding on the behaviour of a material necessitates insight on the shape adopted by the phases. 
Moreover, along with the morphology, the distribution of the phases also contributes to the properties of the material.
Given the influence of the shape and distribution of the phases on the properties, the microstructural evolution in the absence of the phase transformation is extensively studied~\cite{zhao2003microstructural,gogia1998microstructure}.
These investigations, while assisting the formulation of appropriate processing technique that render desired properties, explicate the morphological stability of the microstructure, particularly at high temperatures. 

The morphological evolution of a microstructure is significantly different from the phase transformation.
While the phase change, depending on the nature of the transformation, involves a definite driving force in the form of supersaturation or undercooling, the morphological evolution predominantly ensues a chemical equilibrium~\cite{smallman2016modern}. 
Therefore, the volume fractions of the phases are characteristically preserved during the shape change.
Furthermore, the phase transformations are often steady-state in nature, $i.e,$ the driving force remains constant all-through the evolution~\cite{huang1981overview,glicksman2000diffusion}, except under specific condition~\cite{cahn1963divergent}.
In contrast, the curvature-difference which dictates the shape change in the phases decreases progressively with time~\cite{courtney1989shape}. 
Despite these differences, one critical aspect of the morphological transformation that convolutes its investigations is the demand for the three-dimensional projection of the microstructure~\cite{wang2010quantitative}.
Considerable understanding of a phase change can be gained by analysing the microstructural evolution in two dimensions. 
However, such conventional treatment generally fails to offer a convincing insight on the morphological transformation.
Owing to this reason, theoretical techniques are often adopted to complement the experimental observation, and deepen the understanding of the \lq shape instabilities\rq \thinspace~\cite{cline1971shape,lee1989two}.
In the present work, a well-established numerical approach called phase-field method is employed to analyse a complex form of morphological transformation in polycrystalline systems.

\subsection{Morphological changes}\label{sec:theory}

\begin{figure}
    \centering
      \begin{tabular}{@{}c@{}}
      \includegraphics[width=0.75\textwidth]{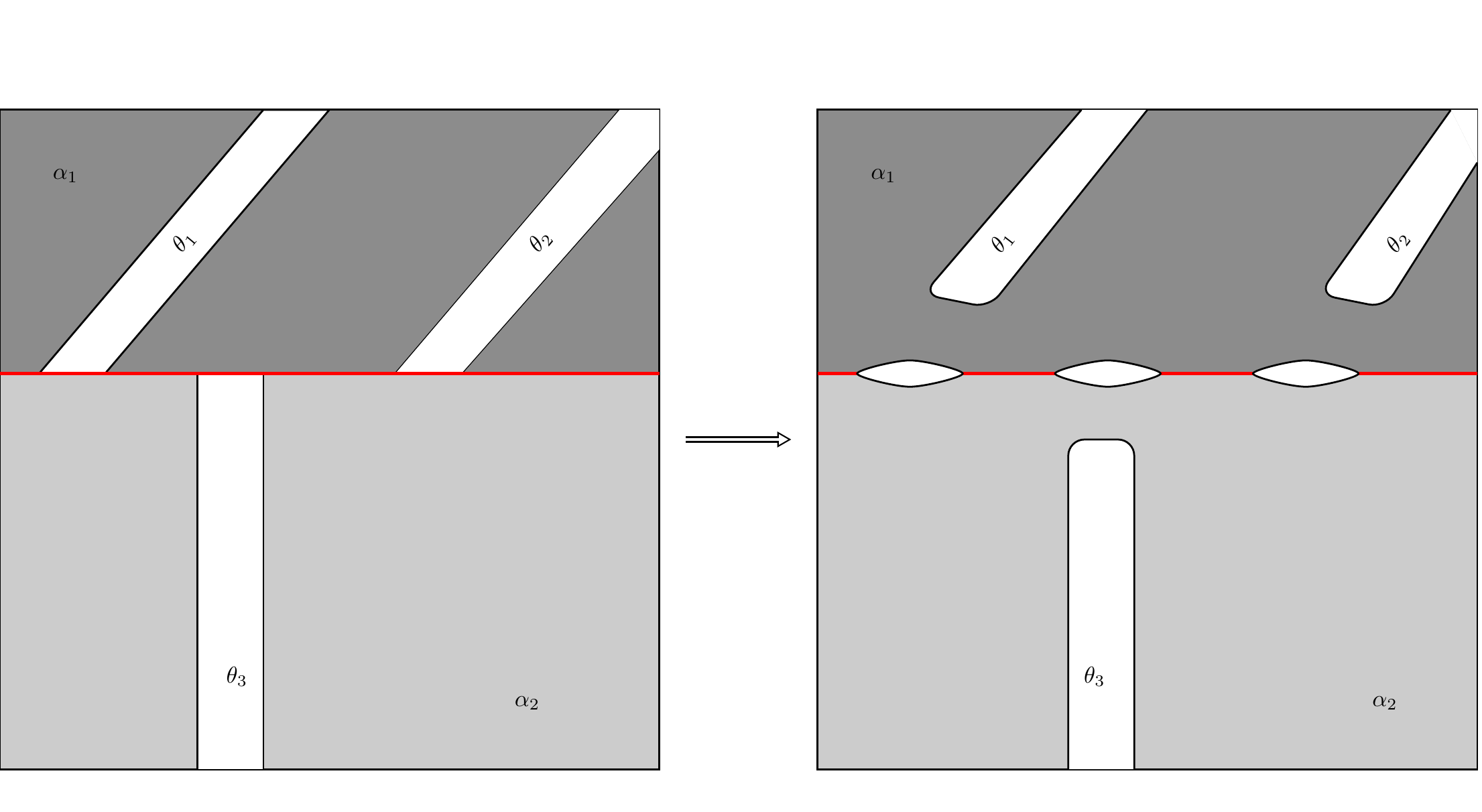}
    \end{tabular}
    \caption{ Schematic representation of grain-boundary assisted ovulation observed during the pearlite spheroidization.
    \label{fig:gb_schematic}}
\end{figure}

The theoretical treatment, which investigates the stability of fluid jet, involves considering an infinitely long structure and studying its response to the imposed perturbation~\cite{rayleigh1879capillary,rayleigh1878instability}.
A fundamental limitation of extending this technique to investigate morphological stability of a microstructure is that the infinitely long structures are rarely observed in a solid-state material.
In fact, it has been identified that at high temperatures, the seemingly continuous shapes break down into finite structures governed by the through-thickness boundaries~\cite{sharma2000instability,zherebtsov2011spheroidization,sandim2004annealing}.
This fragmentation is referred to as boundary splitting.
Fragmentation of the continuous structures plays a vital role in the morphological transformation of a microstructure.
This event introduces termination to a given precipitate, as it converts a seemingly infinite shape to a finite one. 
The subsequent morphological evolution is governed by the curvature-difference introduced by the termination.
In sub-critical annealing, wherein the through-thickness boundaries are purposely introduced through mechanical treatments, the boundary splitting form an integral part of spheroidization~\cite{lupton1972influence,poths2004effect}.
The morphological changes associated with the boundary splitting can be elucidated by considering the well-known shape instability called thermal grooving.

As opposed to boundary splitting, recent experimental observations have unraveled a unique shape instability wherein the precipitates abutting the grain boundary break down in a characteristic fashion~\cite{arruabarrena2014influence,li2016microstructure}.
A schematic representation of this fragmentation process is illustrated in Fig.~\ref{fig:gb_schematic}.
It is observed that the \lq ovulation \rq \thinspace of the precipitates close to the grain boundaries initiates shape transformation accompanying the pearlite spheroidization.
Based on the preliminary investigation, this morphological evolution is attributed to the grain-boundary diffusion.
However, despite its contribution to the spheroidization~\cite{ji2018effect,kim2016improvement}, the grain-boundary assisted fragmentation has not been comprehensively studied yet.
Therefore, in this work, this particular form of shape instability is extensively analysed to elucidate the underlying mechanism.

Since the morphological evolution in Fig.~\ref{fig:gb_schematic} ~involves at least two grains ($\alpha_1$ and $\alpha_2$) and two phases ($\alpha$ and $\theta$), elementary techniques which are confined to an isolated precipitate in a matrix cannot be adopted to examine this shape instability.
Moreover, the transformation includes morphological evolution of the detached precipitate that follows the initial ovulation.
Therefore, the theoretical approach should combine grain-boundary (or) interfacial diffusion, in addition to the volumetric diffusion. 
A phase-field model which fulfills the central requirements of multiphase consideration and encompasses different modes of mass transfer is formulated and employed in the current investigation.

\subsection{Governing thermodynamics}

The morphological changes exhibited by the phases in chemical equilibrium are delineated by considering the influence of curvature on the chemical potential~\cite{johnson1965generalization,cahn1982surface}.
The change introduced by the curvature in the equilibrium chemical potential is expressed by the well known Gibbs-Thomson relation as
\begin{align}\label{eq:gibbs_law}
 \mu(K) = \mu^{0}_{\text{eq}} + \underbrace{\gamma V_m K}_{:=\tilde\mu (K)},
\end{align}
where $\gamma$ and $V_m$ are the interfacial energy and molar volume of the diffusing element~\cite{tian1987kinetics,mullins1957theory}.
Moreover, the constant chemical-potential across the flat surface of the chemically-stable phases is represented by $\mu^{0}_{\text{eq}}$.
The mean curvature $K$ in Eqn.~\eqref{eq:gibbs_law} is the summation of the curvatures along the principal directions, and is written as
\begin{align}\label{eq:curvature}
 K =\frac{1}{2} \left( \frac{1}{R_1}+\frac{1}{R_2} \right),
\end{align}
where $R_1$ and $R_2$ are principal radii of curvature.
In a given system, any disparity in the curvature introduces a gradient in chemical potential which consequently induces migration of atoms.
This flux of atoms establishes the morphological changes, which progressively decrease the overall interfacial energy of the system while reducing the governing curvature-difference.
The velocity of the atoms migrating under the influence of the induced potential-gradient is expressed as
\begin{align}\label{eq:atm_v1}
 v_a= -\frac{D}{RT}\n \mu(K) = -\frac{D}{RT}[\underbrace{\n \mu^{0}_{\text{eq}}}_{=0} + \gamma V_m \n K].
\end{align}
In the above relation and as described in~\cite{jost1952diffusion}, the kinetic constant $D$ is the diffusivity with $R$ and $T$ representing the universal gas constant and temperature, respectively.
Since the chemical potential of all the components is constant across the flat interface in chemical equilibrium, its gradient vanishes in Eqn.~\eqref{eq:atm_v1}.

Based on the velocity of the migrating species, the flux is ascertained by including the concentration which distinguishes the precipitate from the matrix.
Correspondingly, the flux of the atoms is written as
\begin{align}\label{eq:flux1}
 \vv{J}= -\frac{D\gamma V_m}{RT}c_{\text{eq}}^{\delta}\n K,
\end{align}
where $c_{\text{eq}}^{\delta}$ is the equilibrium composition of the precipitate-$\delta$ expressed in mole fraction.
The description of the flux in Eqn.~\eqref{eq:flux1} assumes $c_{\text{eq}}^{\alpha} << c_{\text{eq}}^{\delta}$, where $c_{\text{eq}}^{\alpha}$ is the equilibrium concentration of matrix.
In a system wherein this condition is not satisfied, $c_{\text{eq}}^{\delta}$ in Eqn.~\eqref{eq:flux1} is replaced by $c_{\text{eq}}^{\delta} - c_{\text{eq}}^{\alpha}$.

\begin{figure}
    \centering
      \begin{tabular}{@{}c@{}}
      \includegraphics[width=0.75\textwidth]{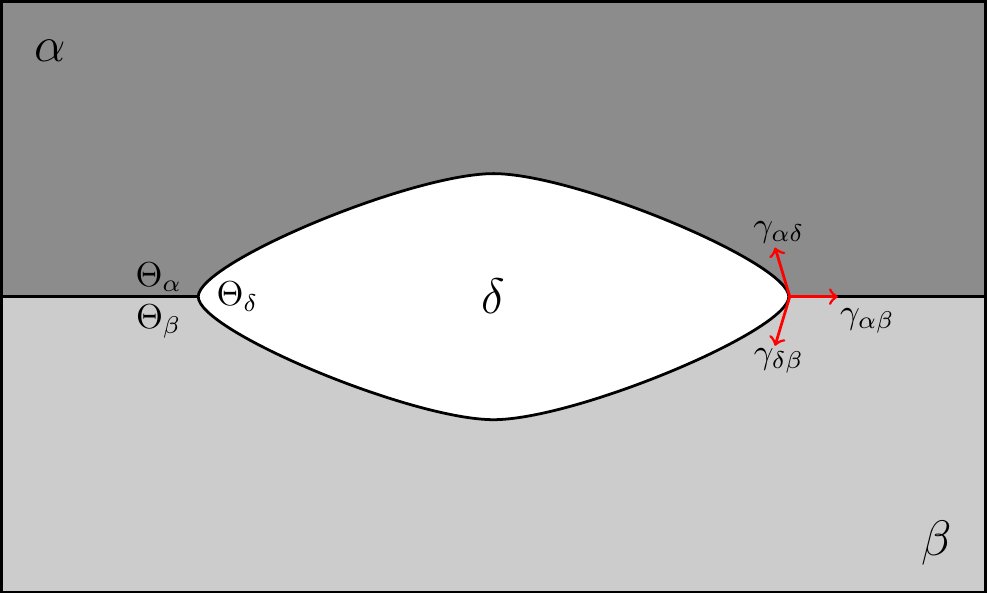}
    \end{tabular}
    \caption{ Equilibrated morphology of the precipitate-$\delta$ governed by the interfacial energies ($\gamma_{\alpha\beta}$, $\gamma_{\alpha\delta}$ and $\gamma_{\delta\beta}$) associated with the triple junction.
    \label{fig:schematic}}
\end{figure}

The overall flux of atoms, whose migration is governed by the gradient in the chemical potential, can be dissociated into tangential and normal component~\cite{srinivasan1973theory}.
While the tangential component is restricted to the flux along the surfaces, the other migration paths are encompassed by the normal flux.
(The interface separating the phases and the grain boundaries are collectively referred to as surface henceforth in this work.)
Correspondingly, the overall atomic flux governing the morphological evolution is expressed as
\begin{align}\label{eq:flux2}
 \vv{J}&=\vv{J_s}+\vv{J_n} \\ \nonumber
 & = -\frac{1}{RT}c_{\text{eq}}^{\delta}\left( \gamma V_m D^{\text{surf}}\n_{s} K + D^{\text{vol}}\n_{n} \tilde\mu (K) \right),
\end{align}
wherein the diffusion coefficient is appropriately distinguished into surface and volume diffusivity, $D^{\text{surf}}$ and $D^{\text{vol}}$, respectively.
In Eqn.~\eqref{eq:flux2}, the surface gradient is represented by $\n_{s}$, while the remaining gradients are included in $\n_{n}$.
In other words, the influence of curvature which is not confined to the surface is encompassed by $\n_{n}\tilde\mu (K)$. 
The temporal evolution of a small surface element $\eta$, dictated by the atomic fluxes, is ascertained by
\begin{align}\label{eq:flux3}
 \frac{\partial \eta}{\partial t} = \frac{V_m}{RT}c_{\text{eq}}^{\delta}\left( \gamma V_m D^{\text{surf}}\vv{\Delta}_s K + D^{\text{vol}}\n_{n} \tilde\mu (K) \right).
\end{align}
Since the migration of the interface under surface diffusion is comprehended by the surface gradient of the corresponding fluxes, the gradient in the curvature is replaced by $\vv{\Delta}_s( = \n_{s}^2)$ in the tangential component of Eqn.~\eqref{eq:flux3}~\cite{weatherly1975stability}.

In a multiphase polycrystalline system, the minimisation of the overall interfacial energy of the system extends beyond the morphological evolution of the individual precipitate.
At triple junctions, the equilibrium condition is established only when the forces introduced by the individual interfaces are zero at that point.
Projecting the interfacial energies, which primarily contribute to the forces at the triple junction, as von Neumann triangle, the Young's balance of force can be expressed as the law of sines~\cite{neumann1894vorlesungen,rowlinson2013molecular}.
Correspondingly, for a schematic distribution of phases shown in Fig.~\ref{fig:schematic}, the morphology of the precipitate-$\delta$ is dictated by the relation
\begin{align}\label{eq:triple_junc} 
\frac{\gamma_{\alpha\beta}}{\sin \Theta_{\delta}} = \frac{\gamma_{\alpha\delta}}{\sin \Theta_{\beta}} = \frac{\gamma_{\delta\beta}}{\sin \Theta_{\alpha}}.
\end{align}
Moreover, it is evident that the above relation is pertinent to the morphological evolution illustrated in Fig.~\ref{fig:gb_schematic}.

\section{Numerical framework}

Phase-field approach is known for its ability to simulate microstructural evolutions~\cite{provatas2011phase}.
The microstructural changes analysed by this numerical technique include both phase transformation and energy-minimising interface evolutions.
While a wide-range of phase changes is quantitatively examined by incorporating appropriate driving force based on the CALPHAD data~\cite{mushongera2018phase,amos2019phase}, the energy-minimising transformations have largely been directed towards grain growth~\cite{perumal2018phase,perumal2017phase}.
Recently, curvature-driven transformations, wherein the volume fractions of the phases are preserved, are increasingly analysed using phase-field approach~\cite{tian2014phase,yang2018phase,chakrabarti2017grain}.
However, as opposed to the conventional Cahn-Hilliard treatment, the volume-preserved morphological evolutions have been studied in the Allen-Cahn framework~\cite{amos2018phase2,amos2018volume,mittnacht2018understanding}.
In the present work, such numerical approach is employed to understand the grain-boundary assisted fragmentation of three-dimensional rods.

\subsection{Multicomponent Multiphase-field model}

Generally, the microstructural changes are comprehended by tracking the temporally-evolving sharp interface~\cite{glimm2001conservative}.
With increasing complexity in the morphology of the evolving phases, the interface tracking becomes numerically more complicated.
Moreover, it is an extremely arduous task to monitor the interface, when the transformation involves singularity events like fragmentation.
In the phase-field approach, this difficulty is circumvented by the introduction of a scalar variable, called phase-field.
The microstructural evolution is, therefore, dictated by the spatio-temporal evolution of the phase-field, instead of the interface migration.

The characteristic inclusion of the phase-field replaces the sharp interface with a diffuse region wherein the scalar variable exhibits a smooth transition~\cite{hohenberg1977theory}.
Therefore, the entire system can be distinguished into bulk phase, where the phase-field assumes a constant value, and diffuse interface which separates the bulk phases.
According to~\cite{Choudhury2012}, the energy density of a system can be expressed as
\begin{align}\label{eq:functional}
  \mathcal{F}(\vphi, \n \vphi, \vc, T) &= \mathcal{F}_\text{intf}(\vphi, \n \vphi) + \mathcal{F}_\text{chem}(\vc, T, \vphi)\\ \nonumber
  &=\int_{V} \epsilon a(\vphi, \n \vphi)+\frac{1}{\epsilon} w(\vphi)+f_\text{chem}(\vc, T, \vphi)\diff V,
\end{align}
where the contribution of the bulk phases and the interfaces are denoted by $\mathcal{F}_\text{chem}$ and $\mathcal{F}_\text{intf}$, respectively.
Owing to the multiphase nature of the formulation, the phase-fields are represented as a N-tuple, $\vphi = \{\phia,\phib,\dots,\phi_{N}\}$ with $N$ denoting the total number of phases.

It is evident from Eqn.~\eqref{eq:functional} that, in a system of volume $V$, the interface contribution comprises of a gradient energy term ($a(\vphi, \n \vphi)$) and penalising potential ($w(\vphi)$)~\cite{nestler2005multicomponent}.
In a multiphase system, the gradient energy, which is formulated based on the gradient of the phase-field in the diffuse region, is written as
\begin{align}\label{eq:a}
  \epsilon a(\vphi, \n \vphi) = \epsilon \sum_{\alpha < \beta}^{N} \gab \underbrace{|{\phi_{\alpha} \bm{\nabla}{\phi}_{\beta}} - {\phi_{\beta} \bm{\nabla}{\phi}_{\alpha}}|^2}_{:=|\qab|^2},
\end{align}
where $\epsilon$ is length-scale parameter that defines the interface width.
In Eqn.~\eqref{eq:a}, the energy density of the interface separating phase$-\alpha$ and $-\beta$ is represented by $\gab$.
The gradient energy expressed in Eqn.~\eqref{eq:a} is limited to the isotropic conditions, but can be easily extended in case of anisotropic features~\cite{tschukin2017concepts}.
However, by appending a prefactor, $\bar{c}(\qab)$, as a function of the gradient vector, anisotropy in the interfacial energy can be introduced~\cite{tschukin2017concepts}. 

The penalising potential ensures that the value of the phase-field is constant at either ends of the diffuse interface. 
Conventionally, this is achieved by a well-type function with its minimas corresponding to the desired phase-field values.
In the present model, however, obstacle-type potential is employed owing to its numerical efficiency~\cite{garcke1999multiphase}.
Accordingly, the penalising function in Eqn.~\eqref{eq:functional} reads
\begin{equation}\label{eq:potential_energy}
 \frac{1}{\epsilon}\omega(\vphi)= 
 \begin{cases}
 \frac{16}{\epsilon\pi^2} \underset{\alpha < \beta}{\sum} \gab \phia \phib + \frac{1}{\epsilon}\underset{\alpha < \beta < \delta}{\sum} \gamma_{\alpha \beta \delta} \phia \phib \phi_{\delta},&  ~\vphi \in \mathcal{G}\\
 \infty &  ~\vphi \notin \mathcal{G},
 \end{cases}
\end{equation}
where the higher order term, $\gamma_{\alpha \beta \delta} \phia \phib \phi_{\delta}$, prevents the formation of the spurious phases.
As expressed in Eqn.~\eqref{eq:potential_energy}, penalising potential operates under a definite condition imposed through the Gibbs simplex,
\begin{equation}\label{eq:gibbs_simplex}
 \mathcal{G}=\left\{ \vv{\phi} \in \mathbb{R}^N: \sum_\alpha \phi_\alpha = 1, \phi_\alpha \geq 0 \right\}.
\end{equation}
This Gibbs simplex cumulatively enhances the efficiency of the obstacle-type potential in the multiphase formulation.

The contribution of the bulk phases to the overall energy density in Eqn.~\eqref{eq:functional} is dictated by the respective composition.
The total free-energy contribution of the phases is expressed as
\begin{align}\label{free_energy}
 f_\text{chem}(\vphi, \vc, T) = \displaystyle\sum_{\alpha}^{N} f_\text{chem}^{\alpha}(\vc^{\alpha}, T) \phia(\vv{x},t),
\end{align}
wherein spatially varying phase-field is employed to interpolate individual free-energy densities.
The free-energy density of the phases is represented by $f_\text{chem}^{\alpha}(\vc^{\alpha}, T)$ in Eqn.~\eqref{free_energy}.
Since a system with more than one independent concentration is considered, akin to the phase-field $\vphi$, the composition is represent as tuple vector,
\begin{align}\label{c_tuple}
 \vc^{\alpha} = \{ c^{\alpha}_{i}, c^{\alpha}_{j}, \cdots,c^{\alpha}_{k}\},
\end{align}
where $k$ represents the number of components, including the solvent (matrix).

It is evident from the energy density formulation in Eqn.~\eqref{eq:functional} that the contributions from the interface and the bulk phases are decoupled.
However, improper treatment of the constitutive variables in the diffuse region separating the phases leads to an unphysical influence of bulk phase on the interface~\cite{wheeler1992phase,tiaden1998multiphase}.
The inefficient decoupling, by adding excess energy to the interface, effects the entire microstructural evolution.
In a system, wherein bulk phases are distinguished by the respective concentration, the energy contributions are decoupled by assuming that the respective phases co-exist in a chemical equilibrium in the diffuse interface~\cite{kim1999phase,eiken2006multiphase}.
Under equilibrium, the chemical potential of a component $i$, in the region separating two phases, can be related as 
\begin{align}\label{eq_chm}
  \frac{\partial f_\text{chem}^{\alpha}(\vc^{\alpha}, T)}{\partial c_{i}} =   \frac{\partial f_\text{chem}^{\beta}(\vc^{\beta}, T)}{\partial c_{i}} = \mu_i(\vv{x}),  \qquad \forall \; \{\alpha,\beta \in [N], i\in[k]\}.
\end{align}
Correspondingly, the phase-dependent concentrations are interpolated as  
\begin{align}\label{c_int}
 \vc = \sum_{\alpha}^{N} \vc^{\alpha} \phia(\vv{x},t),
\end{align}
through spatially varying equilibrated chemical-potential in Eqn.~\eqref{eq_chm}.
This interpolation scheme averts any influence of the bulk phases on the diffuse interface.

For a given interface and bulk contribution, the evolution of the multiphase multicomponent system is directed towards a phenomenological decrease in the overall energy density described in Eqn.~\eqref{eq:functional}. 
Correspondingly, the temporal evolution of the phase-field is formulated as
\begin{align}\label{phase_evolution}
 \epsilon \frac{\partial \phia}{\partial t}=-\frac{1}{\tilde{N}}\sum_{\alpha<\beta}^{1}\frac{1}{\tau_{\alpha\beta}}\left[ \frac{\delta \mathcal{F}(\vphi, \n \vphi, \vc, T)}{\delta \phia}-\frac{\delta \mathcal{F}(\vphi, \n \vphi, \vc, T)}{\delta \phib}\right ],
\end{align}
where $\tau_{\alpha\beta}$ is the relaxation constant between phase$-\alpha$ and $-\beta$, which ensures the stability of the diffuse interface during the microstructural evolution.
In evolution Eqn.~\eqref{phase_evolution}, $\tilde{N}$ is the total number of locally active phases, which are invariably less than $N$.

Including the energy contribution delineated in Eqns.~\eqref{eq:a}, ~\eqref{eq:potential_energy} and ~\eqref{free_energy}, the phase-field evolution reads
\begin{align}\label{phase_evolution2}
\begin{split}
 \epsilon \frac{\partial \phia}{\partial t}  &=-\frac{1}{\tilde{N}}\sum_{\alpha<\beta}^{1}\frac{1}{\tau_{\alpha\beta}}\left\{\epsilon \n \cdot \left [ \frac{\partial a(\vphi, \n \vphi)}{\partial \n \phia}-\frac{\partial a(\vphi, \n \vphi)}{\partial \n \phib} \right] -\epsilon \left [ \frac{\partial a(\vphi, \n \vphi)}{\partial \phia}-\frac{\partial a(\vphi, \n \vphi)}{\partial \phib} \right] \right. \\
   & \left. - \frac{1}{\epsilon} \left[ \frac{\partial \omega(\vphi)}{\partial \phia}- \frac{\partial \omega(\vphi)}{\partial \phib} \right]-\frac{8\sqrt{\phia\phib}}{\pi} \Delta f_\text{chem}^{\alpha}\right\}.
  \end{split}
\end{align}
The chemical driving-force which contributes to the phase-field evolution in Eqn.~\eqref{phase_evolution} is denoted by $\Delta f_\text{chem}^{\alpha}$.
For the bulk contribution in Eqn.~\eqref{free_energy}, this driving-force is expressed as
\begin{align}\label{chem_driv1}
\Delta f_\text{chem}^{\alpha} =  \sum_{\alpha<\beta} ^{N}\left \{ \left [f_\text{chem}^{\alpha}(\vc^{\alpha}, T) - \sum_{i=1}^{k-1}\mu_{i}c_{i}^{\alpha} \right] - \left [f_\text{chem}^{\beta}(\vc^{\beta}, T) - \sum_{i=1}^{k-1}\mu_{i}c_{i}^{\beta} \right] \right \},
\end{align}
where $\mu_{i}$ is the chemical potential of component $i$, which acts as a continuous variable, owing to the thermodynamic basis for the interpolation of the fundamental variable in Eqn.~\eqref{eq_chm}.
It is evident from Eqn.~\eqref{chem_driv1} that the contribution of a phase to the chemical driving-force is the Legendre transform of its free-energy density.
These individual contributions, in the present framework, can be treated as the grand chemical-potential density of the phases, such that
\begin{align}\label{grand_chem}
f_\text{chem}^{\alpha}(\vc^{\alpha}(\vv\mu), T) - \sum_{i=1}^{k-1}\mu_{i}c_{i}^{\alpha} \equiv \Psi_\text{chem}^{\alpha}(\vc^{\alpha} (\vv\mu), T) \qquad \qquad \forall \; \alpha \in [N].
\end{align}
Therefore, the chemical driving-force can be written as
\begin{align}\label{chem_driv2}
\Delta f_\text{chem}^{\alpha} = \sum_{\alpha<\beta} ^{N}\left [\Psi_\text{chem}^{\alpha}(\vc^{\alpha}(\vv\mu), T) - \Psi_\text{chem}^{\beta}(\vc^{\beta}(\vv\mu), T)\right],
\end{align}
which indicates that the phase-field evolution is governed by the difference in the individual grand chemical-potential densities.
Owing to the direct influence of the grand potential density on the transformation, as indicated in Eqn.~\eqref{chem_driv2}, phase-field models have been developed by defining the overall energy of the system based on this thermodynamic parameter~\cite{plapp2011unified,aagesen2018grand}.

% Any microstructural transformation can be viewed as a deviation from the static condition which is characterised by the lack of interface migration.
The driving force governing any microstructural transformation can be viewed as a deviation from the equilibrium condition which renders a static interface.
Correspondingly, the chemical driving-force for phase-field evolution in Eqn.~\eqref{chem_driv2} can written as
\begin{align}\label{chem_driv3}
\Delta f_{\text{chem}}^{\alpha}(\vv\mu) & = \sum_{\alpha<\beta} ^{N}\left [\Psi_{\text{chem}}^{\alpha}(\vc_\text{eq}^{\alpha}(\vv\mu_\text{eq}), T) - \Psi_{\text{chem}}^{\beta}(\vc_\text{eq}^{\beta}(\vv\mu_\text{eq}), T) \right ] \\ \nonumber
& +  \sum_{\alpha<\beta} ^{N} \sum_{i=1}^{k} \left \{ \left [ \left. \frac{\partial \Psi_\text{chem}^{\alpha}(\vc^{\alpha}(\vv\mu), T)}{\partial \mu_{i}}   \right\vert_{\mu_{i:\text{eq}}} - \left. \frac{\partial \Psi_\text{chem}^{\beta}(\vc^{\beta}(\vv\mu), T)}{\partial \mu_{i}}   \right\vert_{\mu_{i:\text{eq}}}\right ] (\mu_{i} - {\mu_{i:\text{eq}}})\right \},
\end{align}
where the first term on the right-hand side of Eqn.~\eqref{chem_driv3} represents the equilibrium, while the deviation is introduced by the second term.
When the phases are in chemical equilibrium, and the evolution is exclusively governed by the curvature ($K$), Eqn.~\eqref{chem_driv3} reads
\begin{align}\label{chem_driv3b}
\Delta f_{\text{chem}}^{\alpha}(K) & = \sum_{\alpha<\beta} ^{N}\left [\Psi_{\text{chem}}^{\alpha}(\vc_\text{eq}^{\alpha}(\vv\mu_\text{eq}^0)) - \Psi_{\text{chem}}^{\beta}(\vc_\text{eq}^{\beta}(\vv\mu_\text{eq}^0)) \right ] \\ \nonumber
& +  \sum_{\alpha<\beta} ^{N} \sum_{i=1}^{k} \left \{ \left [ \left. \frac{\partial \Psi_\text{chem}^{\alpha}(\vc^{\alpha}(\vv\mu))}{\partial \mu_{i}}   \right\vert_{\mu_{i:\text{eq}}} - \left. \frac{\partial \Psi_\text{chem}^{\beta}(\vc^{\beta}(\vv\mu))}{\partial \mu_{i}}   \right\vert_{\mu_{i:\text{eq}}}\right ] (\mu_{i}(K) - {\mu_{i:\text{eq}}^0})\right \},
\end{align}
where $\mu_{i:\text{eq}}^0$ is the equilibrium chemical-potential of component$-i$ across the flat interface.
In the absence of any curvature, the corresponding driving forces negate each other
\begin{align}\label{equal_gcp}
\Psi_{\text{chem}}^{\alpha}(\vc_\text{eq}^{\alpha}(\vv\mu_\text{eq}^0), T) =  \Psi_{\text{chem}}^{\beta}(\vc_\text{eq}^{\beta}(\vv\mu_\text{eq}^0), T) \qquad \forall \; \{\alpha,\beta \in [N]\}.
\end{align}
Moreover, from Eqn.~\eqref{grand_chem}, the derivative of the grand chemical-potential yields
\begin{align}\label{dr_gcp}
 \frac{ \Psi_\text{chem}^{\alpha}(\vc^{\alpha} (\vv\mu), T)}{\partial \mu_{i}}=-c_{i}^{\alpha} \qquad \forall\;\{i\in[k]:\;\alpha\in[N]\}. 
\end{align}
By substituting Eqns.~\eqref{equal_gcp} and ~\eqref{dr_gcp} in Eqn.~\eqref{chem_driv3b}, the driving force dictating the curvature-driven transformations is expressed as
\begin{align}\label{chem_driv4}
\Delta f_{\text{chem}}^{\alpha} = \sum_{\alpha<\beta} ^{N}\sum_{i=1}^{k} \left \{ \left ( c^{\beta}_{i:\text{eq}} - c^{\alpha}_{i:\text{eq}} \right) \underbrace{ \left [\mu_{i}(K) - \mu_{i:\text{eq}}^{0} \right ]}_{:=\tilde{\mu_{i}}(K)} \right \},
\end{align}
where $c^{\alpha}_{i:\text{eq}}$ and $c^{\beta}_{i:\text{eq}}$ are the respective equilibrium concentrations of component $i$ in phase $-\alpha$ and $-\beta$~\cite{amos2020limitations}.

Eqn.~\eqref{chem_driv4} indicates that the only material-based parameter which contributes to the driving-force of the curvature-driven transformations is the difference between the constant equilibrium concentrations ($c^{\beta}_{i:\text{eq}} - c^{\alpha}_{i:\text{eq}}$).
Therefore, consistent with sharp interface relation, Eqn.~\eqref{eq:flux3}, the temporal evolution of the phase-field is effectively governed by the deviation in the chemical potential introduced by the curvature.
The evolution of the dynamic variable, $\mu_{i}(K)$, can be formulated by considering temporal change in the concentration. 

Considering the dependencies of the homogenised concentration variable, its temporal evolution can be written as
\begin{align}\label{conc_ev1}
\frac{\partial c_{i}(\mu_{i}(K),\vphi)}{\partial t}=\left( \frac{\partial c_{i}}{\partial \mu_{i}(K)}\right)_{T,\phia}\frac{\partial \mu_{i}(K)}{\partial t} + \left( \frac{\partial c_{i}}{\partial \phia} \right)_{T,\tilde{\mu_{i}}}\frac{\partial \phia}{\partial t}.
\end{align}
However, in a system wherein the morphological changes are predominantly governed by inherent curvature-difference, a gradient in the chemical potential is introduced, as elucidated in Sec.~\ref{sec:theory}.
This potential gradient induces atomic flux, which migrates from the region of high potential (source) to low potential (sink).
In a polycrystalline system, the flux of atoms is governed by volume and grain-boundary diffusion.
Therefore, the curvature-driven concentration evolution should encompass both volume and surface fluxes.

In the present approach, the surface flux is introduced by appropriately formulating the concentration mobility~\cite{zhang2006phase}.
The concentration mobility which includes surface diffusion is expressed as
\begin{align}\label{mobility}
\vv{M}(\vphi) &= \vv{M}^{\text{vol}}(\vphi) + \vv{M}^{\text{surf}}(\vphi) \\ \nonumber
&=\sum_{\alpha=1}^{N}\vv{D}^{\text{vol}}_{ij:\alpha\alpha}\left ( \frac{\partial c^{\alpha}_{i}}{\partial \mu_{j}}\right )_{T} h_{\alpha}(\vphi) \\ \nonumber
&+\sum_{\alpha}^{N-1} \sum_{\beta}^{N-1} \vv{D}^{\text{surf}}_{ij:\alpha\beta}\left [\left ( \frac{\partial c^{\alpha}_{i}}{\partial \mu_{j}} \right )_{T} h_{\alpha}(\vphi) + \left ( \frac{\partial c^{\beta}_{i}}{\partial \mu_{j}} \right )_{T} h_{\beta}(\vphi) \right]\phia\phib,
\end{align}
where $\vv{D}^{\text{vol}}_{ij:\alpha\alpha}$ and $\vv{D}^{\text{surf}}_{ij:\alpha\beta}$ correspond to volume and surface diffusivities represented as a second-order tensor.
The diffusivities are collectively expressed as a symmetric matrix of  $N \times N$ dimension,
\begin{equation}\label{diff_matrix1}
\vv{D}_{ij} = \vv{D}^{\text{vol}}_{ij:\alpha\alpha} + \vv{D}^{\text{surf}}_{ij:\alpha\beta}=
    \begin{bmatrix}
      \vv{D}^{\text{vol}}_{ij:\alpha\alpha} & \vv{D}^{\text{surf}}_{ij:\alpha\beta} & \cdots & \vv{D}^{\text{surf}}_{ij:\alpha N} \\
        & \vv{D}^{\text{vol}}_{ij:\beta\beta} & \cdots & \vv{D}^{\text{surf}}_{ij:\beta N} \\
        & \multirow{2}{*}{\makebox[0pt]{\text{sym.}}}  & \ddots & \vdots \\
        &   &   & \vv{D}^{\text{vol}}_{ij:NN}
    \end{bmatrix}.
  \end{equation}
In a multicomponent setup, the rate of a microstructural evolution is not predominantly confined to the diffusivities of individual components~\cite{umantsev1993ostwald,philippe2013ostwald}.
Accordingly, the diagonal elements in the diffusivity tensor, in Eqn.~\eqref{diff_matrix1}, are treated as a phase-dependent interdiffusivity matrix which reads
\begin{equation}\label{diff_matrix2}
\vv{D}^{\text{vol}}_{ij:\alpha\alpha}=
    \begin{bmatrix}
      \vv{D}^{\text{vol}}_{11:\alpha\alpha} & \vv{D}^{\text{vol}}_{12:\alpha\alpha} & \cdots & \vv{D}^{\text{vol}}_{1k:\alpha\alpha} \\
        \vv{D}^{\text{vol}}_{21:\alpha\alpha} & \vv{D}^{\text{vol}}_{22:\alpha\alpha} & \cdots & \vv{D}^{\text{vol}}_{2k:\alpha\alpha} \\
        \vdots&  \vdots & \ddots & \vdots \\
        \vv{D}^{\text{vol}}_{k1:\alpha\alpha} & \vv{D}^{\text{vol}}_{k2:\alpha\alpha}  &  \cdots & \vv{D}^{\text{vol}}_{kk:\alpha\alpha} 
    \end{bmatrix}.
  \end{equation}
Although quantitative information on the interdiffusivities are rendered by kinetic databases like MOBFe3, it is realised that the role of interdiffusivities is governed by the composition of the phases~\cite{coates1973diffusional,coates1973precipitate}.
Therefore, by analysing the relative influence of the non-diagonal elements in matrix $\vv{D}^{\text{vol}}_{ij:\alpha\alpha}$ in Eqn.~\eqref{diff_matrix2}, the phase-dependent diffusivity can be efficiently defined. 
Similarly, by formulating the surface diffusivities as a scalar product of the respective volume components, the entire mobility can be described based on the interdiffusivity matrix alone.
Susceptibility matrix, $\frac{\partial c^{\alpha}_{i}}{\partial \mu_{j}}$, which is included in Eqn.~\eqref{conc_ev2}, ensures that the diffusion coefficients are constant in a given phase.
Often, these parameters are incorporated from the kinetic CALPHAD databases.
However, when the energy densities of the individual phases are expressed as a polynomial function, the susceptibility matrix are then the inverse of the second-derivative of the free energy~\cite{amos2018phase1}.

\begin{figure}
    \centering
      \begin{tabular}{@{}c@{}}
      \includegraphics[width=1.0\textwidth]{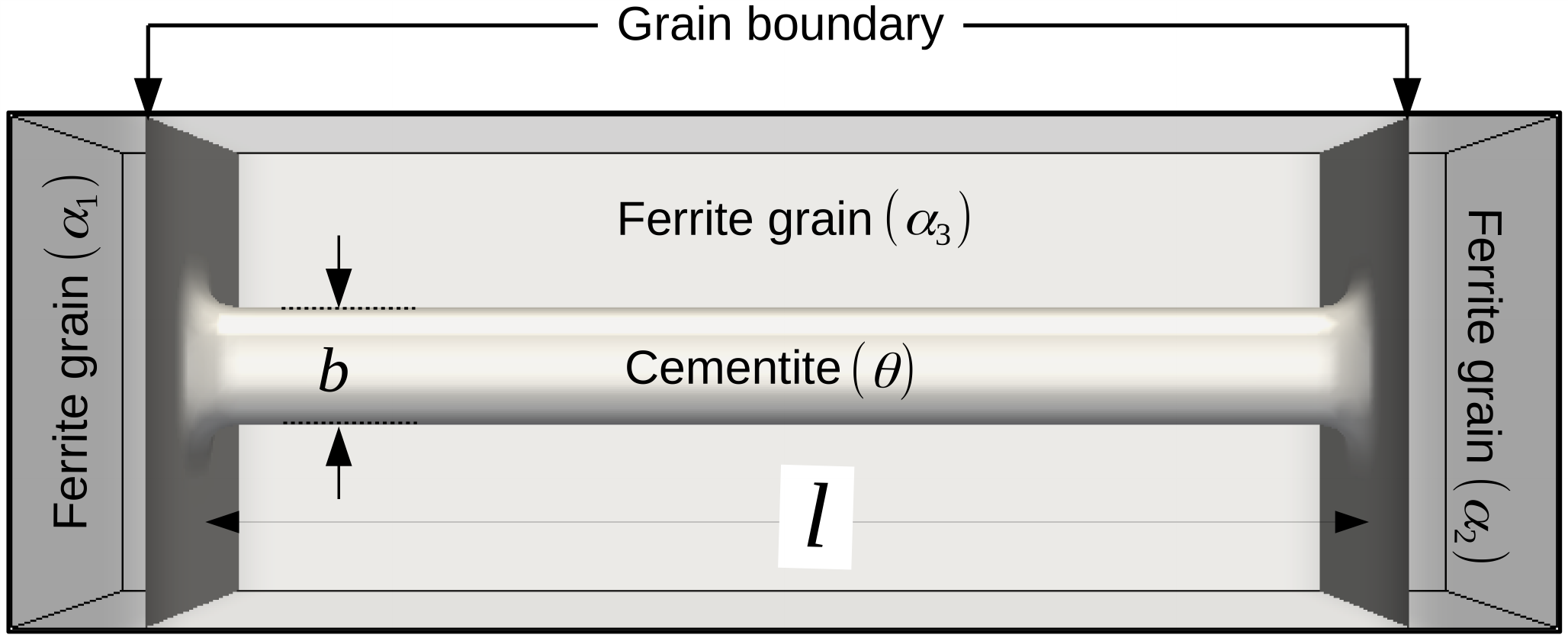}
    \end{tabular}
    \caption{ A representative three-dimensional domain of a multiphase system which includes three ferrite grains ($\alpha_1$, $\alpha_2$ and $\alpha_3$) and a cementite precipitate ($\theta$) of aspect ratio ($\frac{l}{b}=$) 10.
    \label{fig:fig1}}
\end{figure}

By incorporating the mobility in Eqn.~\eqref{mobility}, the concentration evolution under the influence of the curvature is expressed as
\begin{align}\label{conc_ev2}
 \frac{\partial c_{i}(\mu_{i}(K))}{\partial t} & = \n\cdot\left( \sum_{j=1}^{k-1} \vv{M}(\vphi) \n \mu_{j}(K)\right )\\ \nonumber
&=\n\cdot \sum_{j=1}^{k-1} \left( \sum_{\alpha=1}^{N}\vv{D}^{\text{vol}}_{ij:\alpha\alpha}\left ( \frac{\partial c^{\alpha}_{i}}{\partial \mu_{j}}\right )_{T} h_{\alpha}(\vphi) \n \mu_{j}(K)\right) \\ \nonumber
&+ \n\cdot \sum_{j=1}^{k-1} \left(  \sum_{\alpha}^{N-1} \sum_{\beta}^{N-1} \vv{D}^{\text{surf}}_{ij:\alpha\beta}\left [\left ( \frac{\partial c^{\alpha}_{i}}{\partial \mu_{j}} \right )_{T} h_{\alpha}(\vphi) + \left ( \frac{\partial c^{\beta}_{i}}{\partial \mu_{j}} \right )_{T} h_{\beta}(\vphi) \right]\phia\phib  \n \mu_{j}(K)\right).
\end{align}
Substituting the above flux-based temporal evolution in Eqn.~\eqref{conc_ev1}, and re-arranging the terms, the progressive change in the governing chemical potential is written as 
\begin{align}\label{chempot_ev}
\frac{\partial \mu_{i}(K)}{\partial t}=\left \{ \n\cdot\left[ \sum_{j=1}^{k-1} \vv{M}(\vphi) \n \mu_{j}(K)\right ] - \sum_{\alpha=1}^{N}c_{i}^{\alpha}\frac{\partial \phia}{\partial t} \right \} \left [ \sum_{\alpha}^{N} h_{\alpha}(\vphi) \frac{\partial c_{i}^{\alpha}}{\partial \mu_{j}} \right ]_{ij}^{-1}.
\end{align}
As described in Eqns.~\eqref{phase_evolution} and ~\eqref{chem_driv4}, the evolution of the curvature-dependent chemical potential in Eqn.~\eqref{chempot_ev} ultimately dictates the morphological transformation by governing spatio-temporal change in phase-field.

Through asymptotic analysis, it has already been shown that, in a phase-field model wherein the chemical potential acts as the dynamic variable the Gibbs-Thomson relation is effectively recovered despite the introduction of the diffuse interface~\cite{amos2018phase1,KubendranAmos2019_1000095355}.
Moreover, as elucidated in Ref. ~\cite{amos2018chemo}, the interface contribution formulated in Eqns.~\eqref{eq:a} and ~\eqref{eq:potential_energy} establishes equilibrium condition at the triple junction, despite the complex interplay of the driving forces.
%Moreover, for the present interface considerations in Eqns.~\eqref{eq:a} and ~\eqref{eq:potential_energy}, with the incorporation of chemical and elastic bulk driving-forces, it has been shown that the equilibrated condition in Eqn.~\eqref{eq:triple_junc} is established at the triple junctions~\cite{amos2018chemo}.
Owing to this ability of the present numerical approach to recover the sharp interface solutions, it is adopted to investigate the fragmentation of the three-dimensional rod in a multiphase system.

\subsection{Domain configuration}

\begin{figure}
    \centering
      \begin{tabular}{@{}c@{}}
      \includegraphics[width=0.5\textwidth]{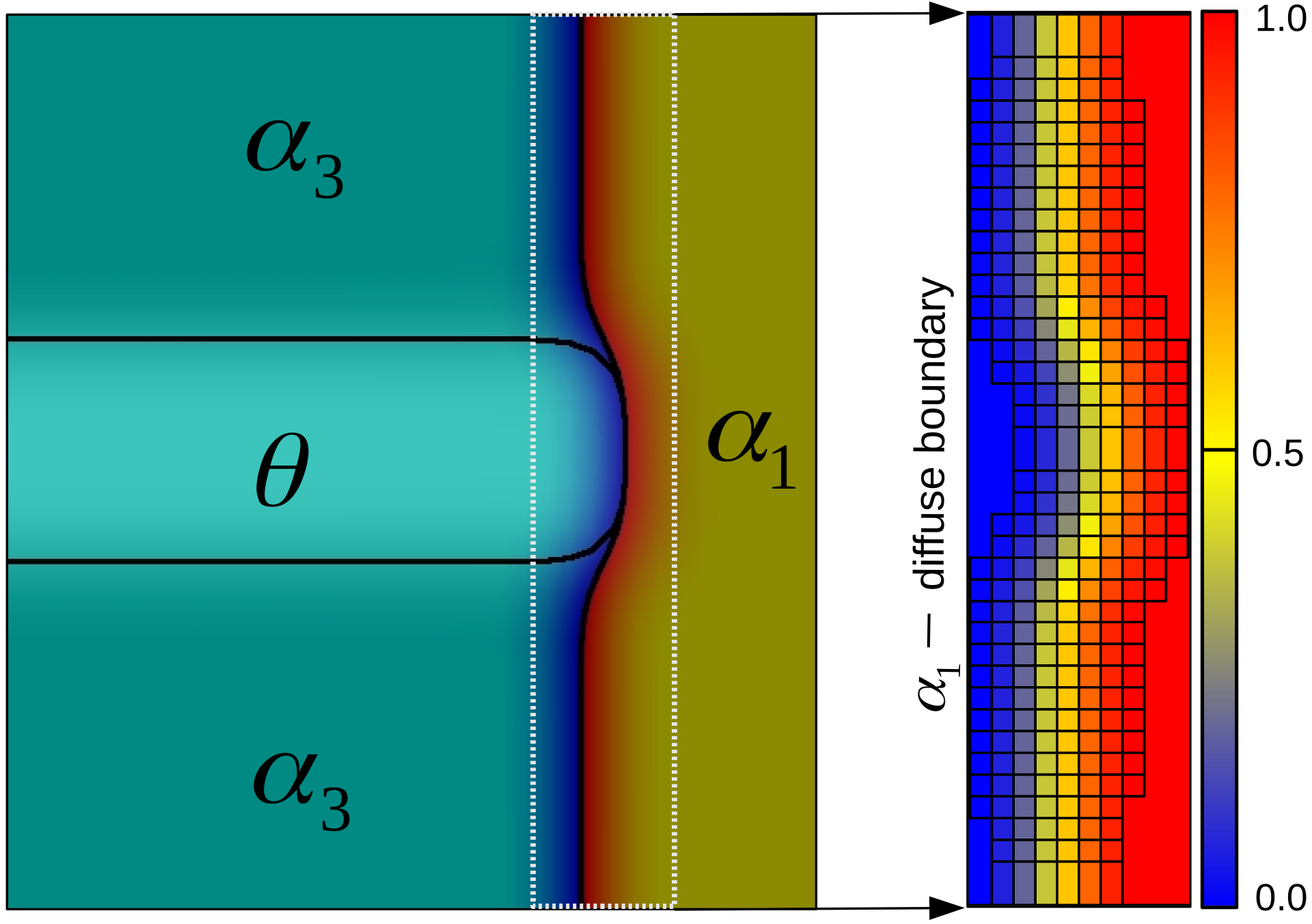}
    \end{tabular}
    \caption{ A cross-sectional representation of the initial configuration of the phases at the grain boundaries wherein the cementite rod is attached. To explicate the condition in which the precipitate is fused with the boundary, the diffuse interface separating the ferrite grains is magnified.
    \label{fig:setup}}
\end{figure}

To elucidate the energy-minimising morphological evolution of the three-dimensional rods in polycrystalline system, representative domains of identical configuration are considered for all simulations in the present analysis.
This elementary multiphase domain is illustrated in Fig.~\ref{fig:fig1}.
Since the form of the shape-instability examined in this work has been observed during the pearlite spheroidization, phases associated with the binary Fe-C system are considered~\cite{arruabarrena2014influence,li2016microstructure}.
Therefore, the free-energy densities of the phases, ferrite and cementite, are defined by incorporating CALPHAD database TCFe8 as elucidated in Ref.~\cite{amos2018phase1}.
Furthermore, equilibrium compositions corroborating mild carbon steels are assigned to avert any phase transformation and establish a chemically stable environment.
In order to distinguish the grains, different phase-fields are assigned to chemically identical phases, $\alpha_1$, $\alpha_2$ and $\alpha_3$. 
The parameters associated with the alloy system are listed in Table.~\eqref{tab:table_1}.
Since the current analysis is confined to the binary Fe-C system, the components of the diffusivity tensor delineated in Eqn.~\eqref{diff_matrix1} adopt a scalar value.
Moreover, it has been identified that morphological evolution in Fe-C systems are governed by volume-diffusion~\cite{tian1987mechanisms,chattopadhyay1977quantitative}.
Accordingly, the concentration mobility is further simplified by assuming marginally different volume and surface diffusivities~\cite{amos2018globularization,amos2018mechanisms}.

\begin{table*}
  \caption{Material parameters involved in the present work.} \label{tab:table_1}
 \centering
 \begin{tabular}{c c c c}
 
  Parameter & Value & Unit \\ [0.5ex]
  \hline
  Temperature ($T$) & 973 & K \\
  Interfacial Energy \\$(\gamma_{\alpha_1\alpha_3}=\gamma_{\alpha_2\alpha_3}=\gamma_{\theta\alpha_i})$ & 0.49 & Jm$^{-2}$\\
  Bulk diffusivity \\($D^{\text{vol}}_{\alpha_1\alpha_1}=\cdots=D^{\text{vol}}_{\theta\theta} = D^{\text{vol}}$) & 2$\times 10^{-9}$ & m$^{2}$s$^{-1}$\\
  Grain boundary diffusivity \\($D^{\text{gb}}_{\alpha_1\alpha_3}=D^{\text{gb}}_{\alpha_2\alpha_3}= D^{\text{gb}}$) & 2.5$\times 10^{-9}$ & m$^{2}$s$^{-1}$\\
  Inter-phase diffusivity \\($D^{\text{intp}}_{\alpha_1\theta}=D^{\text{intp}}_{\alpha_2\theta}=D^{\text{intp}}_{\alpha_3\theta}= D^{\text{intp}}$) & 2.5$\times 10^{-9}$ & m$^{2}$s$^{-1}$\\
  Molar volume ($V_m$) & 7$\times 10^{-6}$  & m$^{3}$/mole \\
  Equilibrium concentration (\text{$c^{\theta}_{\text{eq}}$}) & 0.25 &  mole fraction\\
  Equilibrium concentration (\text{$c^{\alpha}_{\text{eq}}$}) & 0.00067 &  mole fraction\\
  \end{tabular} 
  \end{table*}

The effect of the precipitate size on the transformation mechanism is studied by considering rods of different aspect ratios.
The aspect ratio of a given rod is the ratio of its length ($l$) to its diameter ($d$).
While the diameter of the cementite structure is fixed at 0.02 \textmugreek m, the length is appropriately varied to achieve the desired aspect ratio.
The domain size is suitably devised to avoid any influence of the boundary conditions on the morphological evolution of the precipitate.
However, it is varied depending on the size of the precipitate.

Nucleation of a phase which initiates the microstructural changes during the phase transformation is largely confined to the grain boundaries~\cite{russell1980nucleation}.
When the evolving phase encounters another grain boundary in its path, the characteristic growth is terminated. 
Therefore, in a polycrystalline system, the morphology of the precipitate includes the sections of the grain boundaries to which it is attached.
This effective interlocking of the phases and the respective grain boundaries are imperative for the evolution illustrated in Fig.~\ref{fig:gb_schematic}. 
Lack of such unification of the precipitate terminations with the grain boundaries results in a pre-matured detachment of the rods, thereby completely nullifying the polycrystalline nature of the system.
To ensure that the sections of the grain boundaries abutting the cementite rod form an integral part of its longitudinal terminations, the phase distribution is allowing the peripheral segments of the diffuse interface to penetrate into the neighbouring grains.
This configuration which renders the desired interaction between the precipitate and the corresponding grain boundary is shown in Fig.~\ref{fig:setup}.
The minimal penetration of the cementite boundary into the neighbouring grain is affirmed by imposing the condition 
\begin{align}\label{v_cond}
V_{\theta}^{\alpha_1,\alpha_2}=0,
\end{align}
where $V_{\theta}^{\alpha_1,\alpha_2}$ is the volume fraction of cementite in ferrite grains $\alpha_1$ and $\alpha_2$.
The criterion in Eqn.~\eqref{v_cond} ensures that the unique domain configuration does not influence the resulting morphological transformation.

\subsection{Simulation setup}

The three-dimensional domain considered in the current analysis is discretised using finite-difference technique. 
Besides, the entire domain is uniformly decomposed into identical grids of dimension $\Delta$x=$\Delta$y=$\Delta$z=$1\times 10^{-9}$m, where
the equivalent widths of the cross-section are fixed at 0.08 \textmugreek m $\times$ 0.08 \textmugreek m, while the length is varied to accommodate the size of the cementite rod.

The phase-field and curvature-dependent chemical potential evolutions, Eqns.~\eqref{phase_evolution} and ~\eqref{chempot_ev}, are respectively solved by a finite-difference algorithm operating in an explicit forward-marching Euler's scheme.
Periodic condition is employed along all the boundaries of the three-dimensional domain.
The length parameter which governs the width of the diffuse interface is fixed at $\epsilon=2.5\Delta$x for all the simulation.
The relation between the constant $\epsilon$ and the interface width, along with the other interface properties, emerging from the present formulation, are discussed in the Appendix.
The computational resources are optimally expended through Message Passing Interface (MPI) which elegantly decomposes the three-dimensional domain.

\section{Results and discussion}

\begin{figure}
    \centering
      \begin{tabular}{@{}c@{}}
      \includegraphics[width=1.0\textwidth]{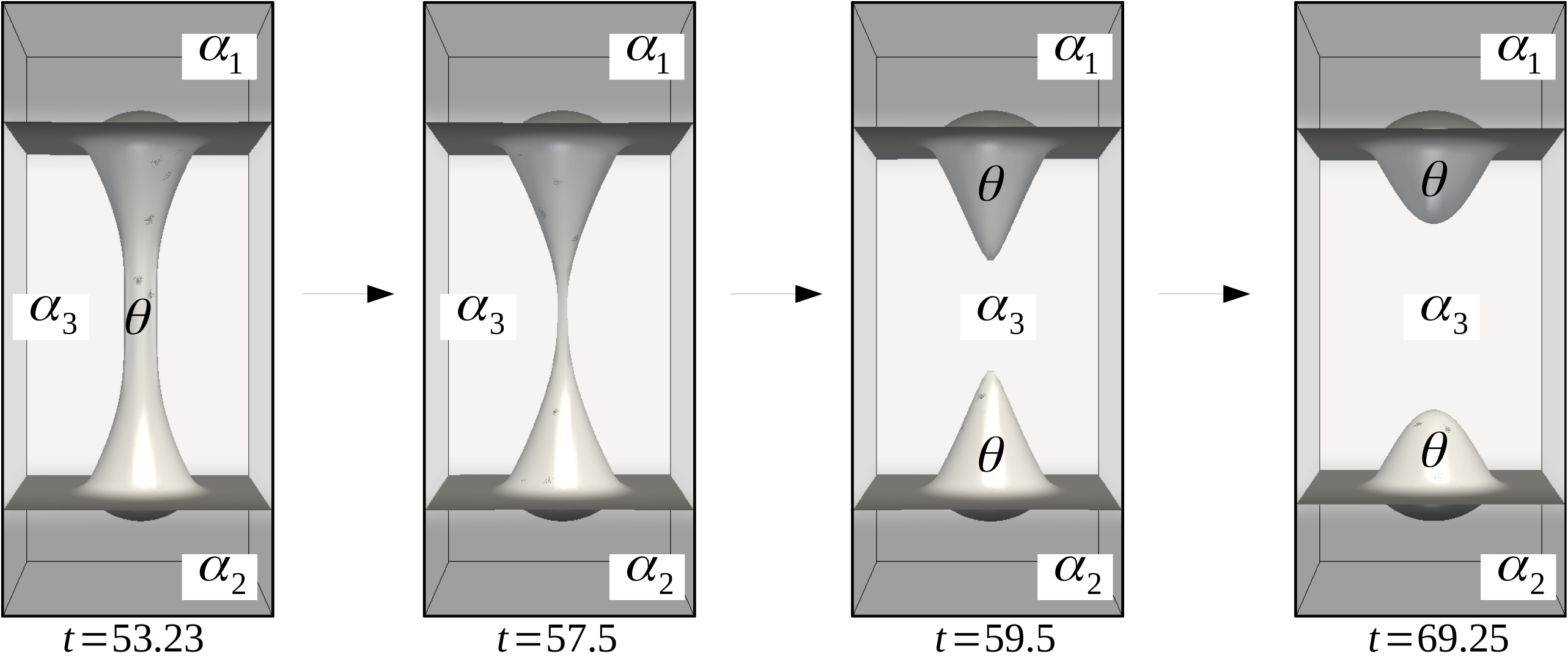}
    \end{tabular}
    \caption{ Shape change exhibited by the cementite rod of aspect ratio 7 pinned to the ferrite grain boundaries.
    \label{fig:fig2}}
\end{figure}

Upon initialising the system, by assigning equilibrium composition, the system is allowed to evolve governed by the minimisation of the overall energy density.
The relaxation constant between the grains and phases is incorporated as a symmetric $4 \times 4$ matrix,
\begin{equation}\label{diff_matrix1_1}
\vv{\tau}=
    \begin{bmatrix}
      \tau_{\alpha_1\alpha_1} & \tau_{\alpha_1\alpha_2} & \tau_{\alpha_1\alpha_3} & \tau_{\alpha_1\theta} \\
        & \tau_{\alpha_2\alpha_2} & \tau_{\alpha_2\alpha_3} & \tau_{\alpha_2\theta} \\
        & \multirow{2}{*}{\makebox[0pt]{\text{sym.}}}  & \tau_{\alpha_3\alpha_3} & \tau_{\alpha_2\theta} \\
        &   &   & \tau_{\theta\theta} 
    \end{bmatrix},
    \end{equation}
with diagonal entities assuming no significance.
The relaxation parameter across the ferrite and cementite phases is determined by 
\begin{align}\label{tau}
\tau_{\alpha_1\theta}=\tau_{\alpha_2\theta}=\tau_{\alpha_3\theta}=0.22\epsilon\frac{\left[c_{\text{eq}}^{\theta}(\mu_{\text{eq}},T)-c_{\text{eq}}^{\alpha}(\mu_{\text{eq}},T)\right]^2}{D_{c} T \left( \frac{\partial c}{\partial \mu}\right)} ,
\end{align}
where $c_{\text{eq}}^{\theta}(\mu_{\text{eq}},T)$ and $c_{\text{eq}}^{\alpha}(\mu_{\text{eq}},T)$ denote the equilibrium carbon-composition of the respective phases~\cite{amos2018mechanisms}.
In the absence of chemical driving-force, $\tau_{\alpha_i \alpha_j}$ is interpreted as the reciprocal of the phase-field mobility~\cite{amos2018chemo}.
Since the ferrite grains in the multiphase domain are exclusively distinguished by the scalar phase-field variables, extremely high values are assigned to $\tau_{\alpha_1\alpha_2}$, $\tau_{\alpha_1\alpha_3}$ and $\tau_{\alpha_2\alpha_3}$, to avert any unphysical migration of the flat grain boundaries.

\subsection{Morphological stability of smaller rods}

A three-dimensional domain akin to Fig.~\ref{fig:fig1} is devised to encompass cementite rod of aspect ratio $7$.
The progressive change in the morphology of these rods in the representative polycrystalline setup is illustrated in Fig.~\ref{fig:fig2}.
In this illustration, $t$ is a dimensionless parameter which is normalised by the constant $\chi=\frac{b^3RT}{D_{\alpha}\gamma_{\alpha_{3}\theta}V_{m}^{2} c_{\text{eq}}^{\alpha}}$, in conformity to the existing works~\cite{courtney1989shape,zherebtsov2011spheroidization}.

As shown in Fig.~\ref{fig:fig2} at $t=53.23$, it is evident that the morphological transformation initiates with a noticeable change in the shape of the cementite terminations which are pinned to the grain boundaries.
In an isolated finite structure, the onset of shape change at the longitudinal ends is seemingly conceivable, since the entire evolution is dictated by the curvature-difference induced by the terminations~\cite{nichols1976spheroidization,amos2018phase2}.
Therefore, smooth ridges are formed by the mass transferred from the receding edges to the neighbouring flat surfaces.
In Fig.~\ref{fig:fig2}, owing to the polycrystalline nature of the system, the precipitate assumes a characteristic morphology at the terminations which is noticeably different from the ridges in the isolated structures.
Moreover, the shape change at the longitudinal ends involves penetration of the cementite into the neighbouring grains.
In other words, fraction of the precipitate, which was originally confined to the parent grain-$\alpha_3$, penetrates into the neighbouring ferrite grains.
The characteristic morphology assumed by the precipitate, which extends beyond a single ferrite grain, indicates that the shape change is initialised by the interaction of the interfacial energies at the triple junctions.

The characteristic morphology established at the longitudinal ends of the rods introduces a curvature-difference.
Governed by the disparity in the curvature, the mass from the remnant sections of the rods, which are confined to the ferrite grains$-\alpha_3$, get deposited in the terminations.
As shown in Fig.~\ref{fig:fig2} at $t=57.5$, the flux results in a significant decrease in the thickness of the rod, particularly in the midriff region, thereby unravelling the high-curvature source.
This mode of the mass transfer is akin to the contra-diffusion observed in the isolated rods wherein the atomic flux from the central region gets deposited in the receding edges. The aforementioned mode is in complete opposition to the regular flow which results in the termination migration.
Ultimately, the progressive deposition of the mass at the longitudinal terminations, as observed at $t=59.5$, results in the fragmentation of the continuous precipitate.
Moreover, as shown in Fig.~\ref{fig:fig2}, the pear-shaped structures resulting from the fragmentation are interlocked with the grain boundaries.
This breaking-up of the continuous structure by the unhindered curvature-driven mass transfer  is referred to as the ovulation.
In isolated rods, it is identified that, in a volume-diffusion governed transformation, the maximum aspect ratio beyond which ovulation sets-in, called critical aspect-ratio, is 8~\cite{mclean1973kinetics}.
However, since the underpinning driving-force for the morphological transformation of the precipitate in a multiphase system is substantially different, the fragmentation is observed in the smaller rods as well.
Following the pinch-off, as illustrated in Fig.~\ref{fig:fig2} at $t=69.25$, the detached precipitates continue to exhibit shape-change to achieve an equilibrated configuration dictated by the relation in Eqn.~\eqref{eq:triple_junc}.

\subsection{Morphological stability of larger rods}\label{sec:10}

\begin{figure}
    \centering
      \begin{tabular}{@{}c@{}}
      \includegraphics[width=0.75\textwidth]{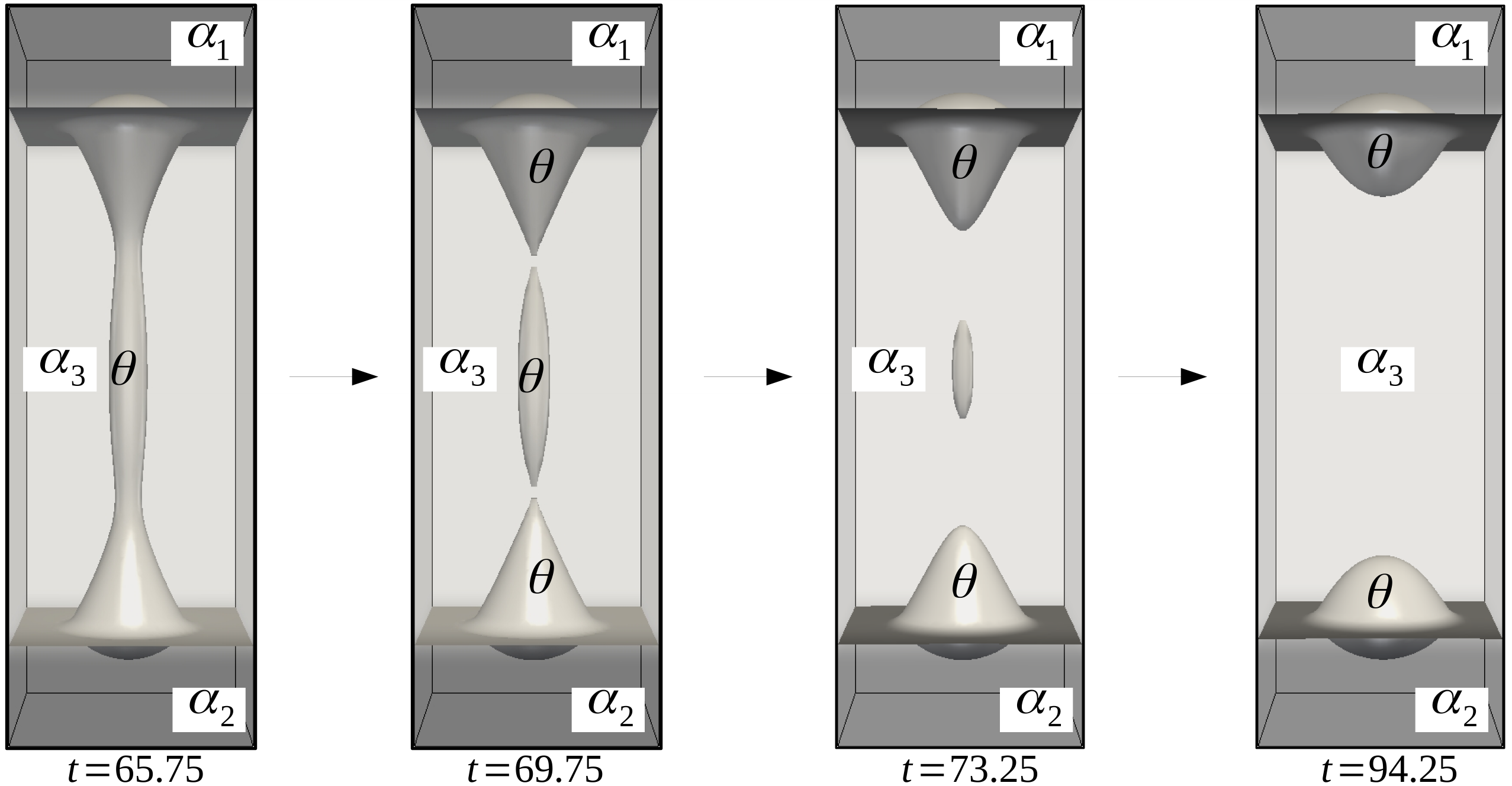}
    \end{tabular}
    \caption{ The energy-minimising morphological transformation of cementite rod of aspect ratio 10 in a polycrystalline setup.
    \label{fig:fig4}}
\end{figure}

In a finite structures, a shift in the transformation mechanism accompanies the increase in the size~\cite{nichols1976spheroidization,amos2018phase2}.
Therefore, to unravel the influence of primary grain-size, thereby the aspect ratio of the rod, on the mechanism of shape change, the morphological evolution in larger domains is analysed.
Fig.~\ref{fig:fig4} illustrates the mode of instability exhibited by the cementite rod of aspect ratio 10 in a multiphase configuration.
Evidently, the transformation mechanism is visibly different from the smaller rods shown in Fig.~\ref{fig:fig2}.

Although the overall configuration of the domain remains unaltered with increase in the size of the cementite rod, the dimension of the remnant section confined to the primary corresponding increases.
Therefore, the shift in the transformation mechanism is predominantly dictated by this increase in the size of the remnant flat-region of the rod which acts as the source of mass transfer.

The energy-minimising event at the triple junctions of the terminations establish a characteristic morphology.
Accordingly, sections of the precipitate which initially confined to the grain$-\alpha_3$ penetrate into the adjacent grains.
Since the volume of cementite is conserved, owing the chemical equilibrium established across the phases, this diffusion of precipitate into ferrite grain$-\alpha_1$ and $-\alpha_2$ is achieved by the curvature-driven mass transfer.
In the smaller rods, the principal source for the atomic flux coincides with the central region of the precipitate.
Therefore, as shown in Fig.~\ref{fig:fig2}, with the progressive mass transfer towards the termination, the thickness of the cementite at the midriff begins to decrease, ultimately leading to the pinch-off at the midpoint.
However, this mode of evolution changes with increase in the size of the rod.
In Fig.~\ref{fig:fig4} at $t=65.75$, the principal sources for the mass transfer can be realised from the thickness of the remnant sections of the rod in grain$-\alpha_3$.
It is clear from the illustration that, with increase in the aspect ratio, the primary source shifts from the central region to the foot of the modified termination.
Correspondingly, separate sources are induced close to each terminations.
The continued mass transfer from the source to the terminations, consequently leads to the simultaneous pinch-off events, as shown in Fig.~\ref{fig:fig4} at $t=69.75$.

The shift in the ovulation site from the midriff region to the foot of the termination increases the number of the fragmentation events.
Additionally, during the pinch-offs, the region between the principal mass-transfer sources detaches from the continuous cementite rod and form an isolated entity which is referred to as the satellite particle.  
Owing to the substantial difference in the volume of the satellite particle and entangled precipitates at the grain boundaries, Ostwald ripening sets-in, as observed at $t=73.25$.
Therefore, the grain-boundary structures coarsen at the expense of the satellite particle.
Ultimately, as shown in Fig.~\ref{fig:fig4} at $t=94.25$, a distribution akin to the evolution of the smaller rod is established with the complete disappearance of the satellite cementite.

\subsection{Size-independent initial morphological changes}

\begin{figure}
    \centering
      \begin{tabular}{@{}c@{}}
      \includegraphics[width=0.4\textwidth]{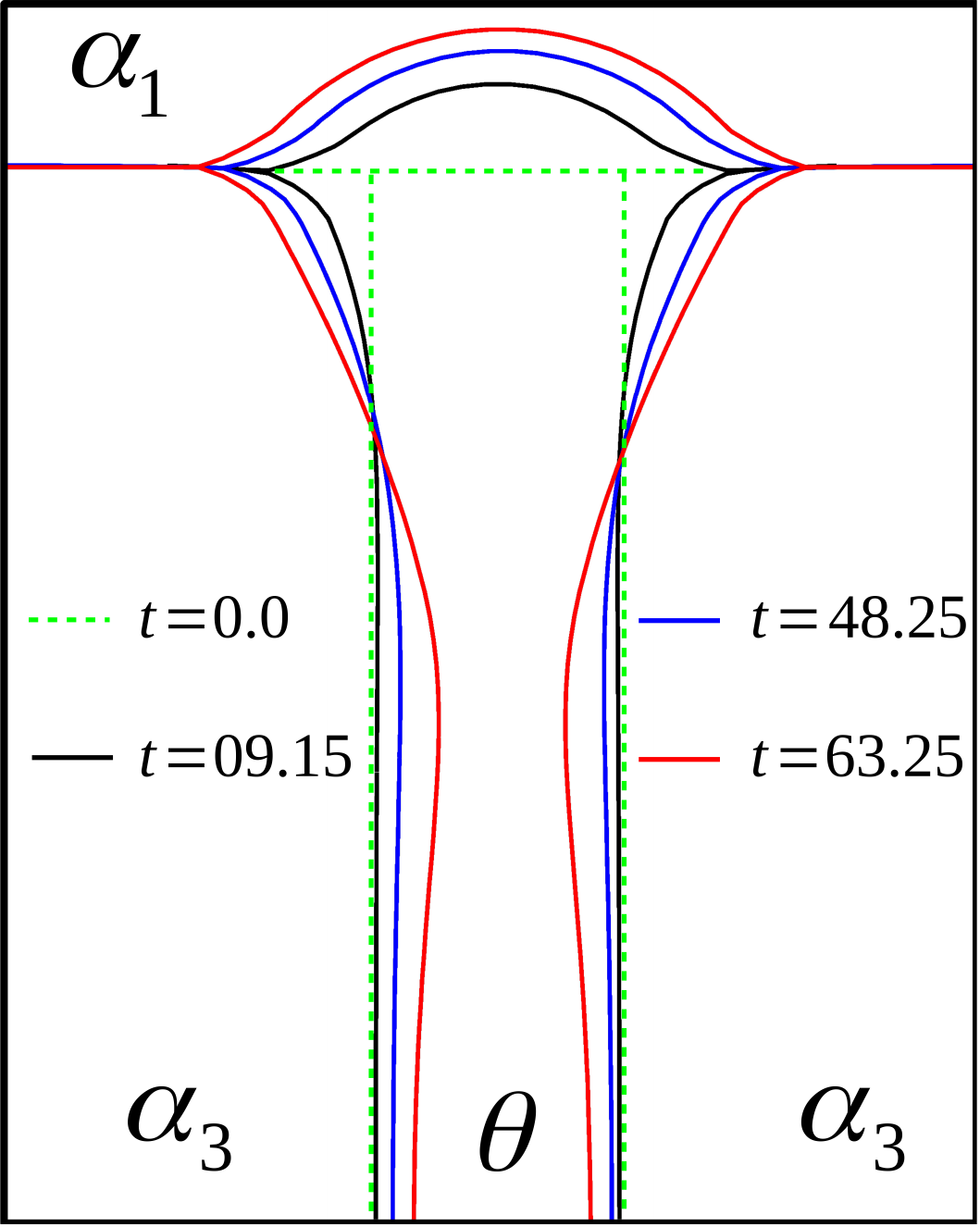}
    \end{tabular}
    \caption{ The isoline representation of the morphological changes in the longitudinal edges of the cementite which establishes the characteristic morphology.
    \label{fig:fig3}}
\end{figure}

Irrespective of the size of the cementite rod, the morphological transformation initiates with the equilibration of the interfacial energies at the triple junction, establishing a characteristic shape at the longitudinal edges of the precipitate.
As shown in Fig.~\ref{fig:fig2} at $t=52.23$ and Fig.~\ref{fig:fig4} at $t=65.75$, the shapes established at the rod terminations by the energy-minimising events at the triple junction are similar, with sections of the cementite percolating in neighbouring grains.
Moreover, it is interesting to note that, both in smaller and larger rods, the time taken for the significant change in the precipitate shape, which includes the ovulation, is noticeably smaller when compared to the entire evolution.
For instance, in the transformation of the cementite rod of aspect ratio $7$, the salient changes occur over a short period of time, $i.e.$ from $t=53.23$ to $t=69.25$, while the temporal evolution leading up to the singularity event is gradual.
Owing to the significance of the morphological changes at the initial stages of the evolution, which are size-independent and predominantly dictate the kinetics, it is independently examined. 

To unravel the temporal evolution which initialises the shape-instability of the rod in polycrystalline setup, the longitudinal cross-section of the domain encompassing precipitate of aspect ratio 10 is considered.
Owing to the two-fold symmetry of the setup, only the upper half of the domain is considered for the representation.
In Fig.~\ref{fig:fig3}, the transformation at the early stages is illustrated through the isoline representation of the interface contours defined by $\Gamma=\{ \vv{x}\in V| \vphi(\vv{x},t)=0.5\}$.

The morphological changes, as mentioned earlier, begin with the change in the angles at the triple junctions.
This is achieved by the equilibration of the tangential forces, predominantly interfacial energies, acting at the triple point.
As shown in Fig.~\ref{fig:fig3} at $t=9.15$, the energy-minimising evolution is accompanied by a characteristic shape-change at the terminations which allows the penetration of the cementite into the adjacent grain.
Furthermore, the marginal and uniform decrease in the thickness of the remnant section of the rod associated with the primary grain, indicates that the source of mass transfer which enables the equilibration and the corresponding morphological evolution is not confined to a definite region.

While the termination grows with time, as illustrated at $t=48.25$ in Fig.~\ref{fig:fig3}, the characteristic shape is retained.
Interestingly, with increase in size at the longitudinal ends, the thickness of the remnant section decreases uniformly.
During contra-diffusion in finite structures, the source is confined to a specific region of the precipitate which exhibits progressive decrease in thickness.
However, the homogeneous thickness of the remnant section unravels that the growth is governed by the mass transfer from the entire flat regions of the cementite rod.
The gradual rate of initial evolution can be attributed to this unique mode of mass transfer, wherein the source is not restricted to a specific section of the structure.
The end of this early transformation is marked by the onset of necking at the foot of the termination at $t=63.25$ in Fig.~\ref{fig:fig3}.
The necking which explicates an increased mass transfer from a specific region turns into the site for ovulation illustrated in Fig.~\ref{fig:fig4}.

\subsection{Time-invariant morphology of the detached structures}

\begin{figure}
    \centering
      \begin{tabular}{@{}c@{}}
      \includegraphics[width=0.75\textwidth]{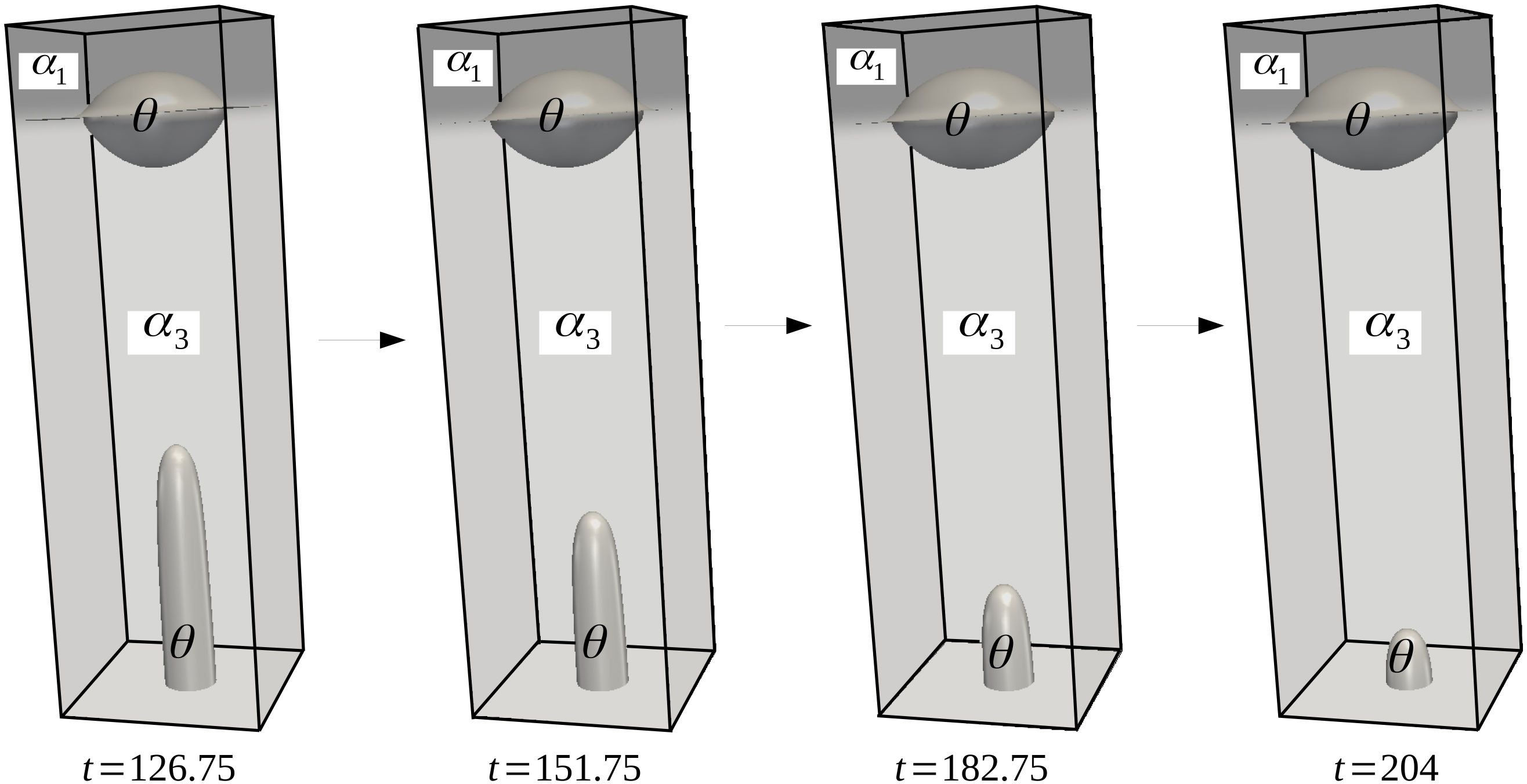}
    \end{tabular}
    \caption{ The transformation following the fragmentation of a continuous rod of aspect ratio 20 in polycrystalline setup. 
    \label{fig:fig5}}
\end{figure}

Since the ovulation and the morphological changes leading to this definitive transformation have been extensively analysed in the previous section, to understand the evolution subsequent to the fragmentation, the stability of the cementite rod of aspect ratio 20 is examined.
In Fig.~\ref{fig:fig5}, the temporal change in the distribution of cementite entities after the pinch-off is illustrated.

As realised in Sec.\ref{sec:10}, with increase in the size of the precipitate, the number of fragmentation events increases and the responsible sources of mass transfer migrate away from the central region.
The shift in the ovulation site from the midriff to the foot of the characteristic shape established at the termination, increases with the aspect ratio of the rods.
Therefore, in the cementite rod of aspect ratio 20, to location of the pinch-off site close to the termination yields a considerably large satellite particle.
The satellite particle resulting from the simultaneous ovulation at two distinct sites is confined to the primary grains$-\alpha_3$.
Despite the increased size, the satellite particle begins to shrink due to the coarsening of the cementite entities which are entangled in the grain boundary.
The gradual shrinking of the isolated precipitate structure, governed by the Ostwald ripening, is illustrated in Fig.~\ref{fig:fig5}.
This coarsening of the grain-boundary particles forms an integral part of the morphological transformation accompanying the shape-instability of large cementite rods in multiphase consideration.

The entire evolution associated with the stability of the larger rods in a polycrystalline system can be elucidated as a successive occurrence of  three distinct events.
The early shape-change involves energy-minimising transformation at the triple junction which renders a characteristic shape to the longitudinal edges pinned to the grain boundaries.
The mass transfer establishing the morphological configuration, at the initial stages, is predominantly governed by the surface diffusion.
Subsequently, the terminals begin to grow, and ultimately, pinch-off, by the atomic fluxes migrating from the remnant section of the rod to the longitudinal ends.
Both surface and volume diffusion contribute to the mass transfer which dictates this morphological transformation.
The coarsening of the termination particle at the grain boundaries, which is shown in Fig.~\ref{fig:fig5}, subsequently follows the fragmentation due to the significant difference in the volume of the resulting entities.
The morphological changes rendered by Ostwald ripening are substantially directed by volume-diffusion through the primary grains$-\alpha_3$. 

\begin{figure}
    \centering
      \begin{tabular}{@{}c@{}}
      \includegraphics[width=0.75\textwidth]{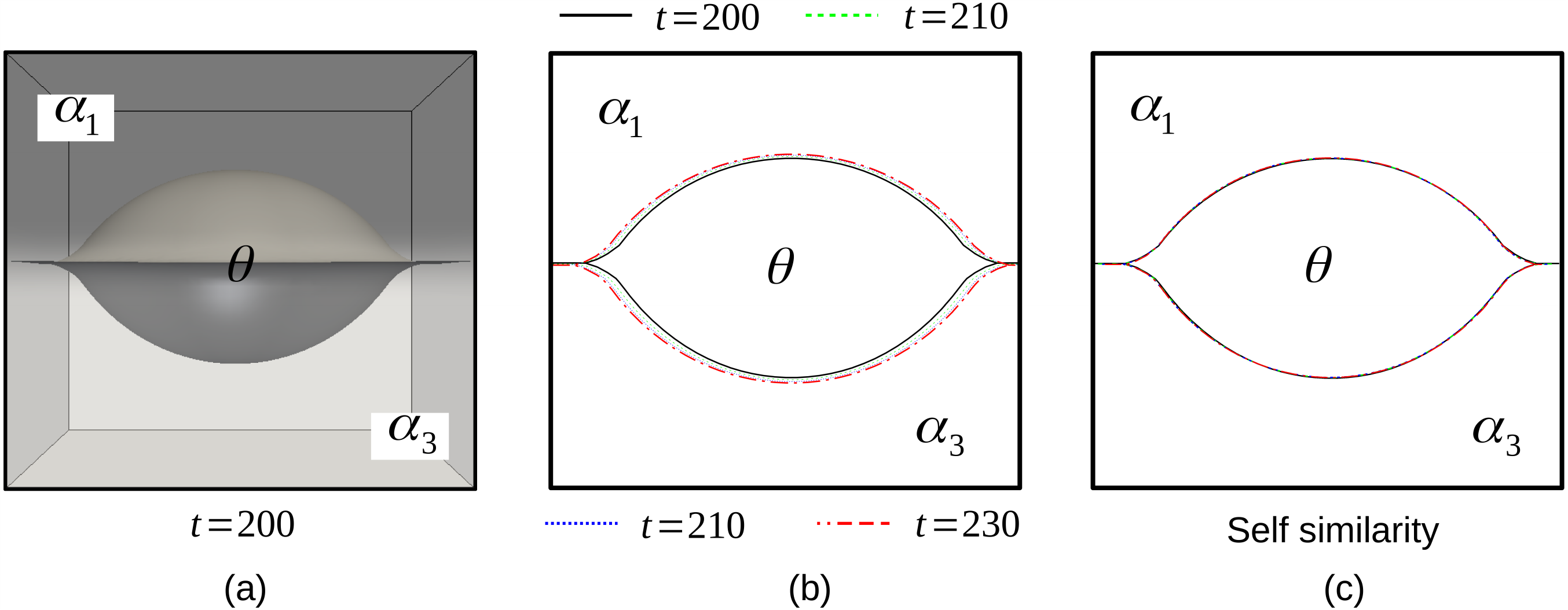}
    \end{tabular}
    \caption{ a) The morphology of the cementite particle detached from the rod of aspect ratio 20 and fastened to the grain boundary at $t=200$. b) Isoline representation of the coarsening of the termination entity at the expense of the satellite particle. c) The time-invariant shape exhibited by the coarsening precipitate unraveled by scaling the dimension.
    \label{fig:fig7}}
\end{figure}

In Fig.~\ref{fig:fig5}, it is noticeable that, despite the dynamic evolution of the fragmented precipitates, the morphology of the cementite particle entangled at the grain boundary remains seemingly unaltered.
This isolated precipitated spanning across the ferrite grains$-\alpha_1$ and $-\alpha_3$ is shown in Fig.~\ref{fig:fig7}a.
Evidently, the shape results from the definite angle subscribed along the radial rim of the precipitate which is fused with the grain boundary.
Since the angle is dictated by the equilibration at the triple point, the unique morphology of the cementite structure is due to interaction between the phase and grain boundary.
In other words, the complex morphological configuration in Fig.~\ref{fig:fig7}a ensures the minimal energy state in a polycrystalline system.

In order to convincingly explicate the shape of the termination precipitate during the coarsening, its evolution is monitored.
The isoline representation of the interface contours pertaining to the cross-section of the domain in Fig.~\ref{fig:fig7}a is adopted to illustrate the growth of the grain-boundary cementite.
The corresponding illustration in Fig.~\ref{fig:fig7}b indicates that, even though the precipitate coarsens, the shape is largely unchanged with time.
However, to unambiguously affirm the time-invariant self similar nature of the cementite morphology, the interface contours at different time-steps are resized and superimposed in Fig.~\ref{fig:fig7}c.
The absolute overlap of the interface contours substantiates the shape-preserved coarsening of the grain-boundary precipitate.
Therefore, despite the deposition of mass from the satellite particle, the characteristic shape dictated by the triple junction remains unperturbed all-through the transformation.

\subsection{Kinetics of the morphological evolution}

\begin{figure}
    \centering
      \begin{tabular}{@{}c@{}}
      \includegraphics[width=0.75\textwidth]{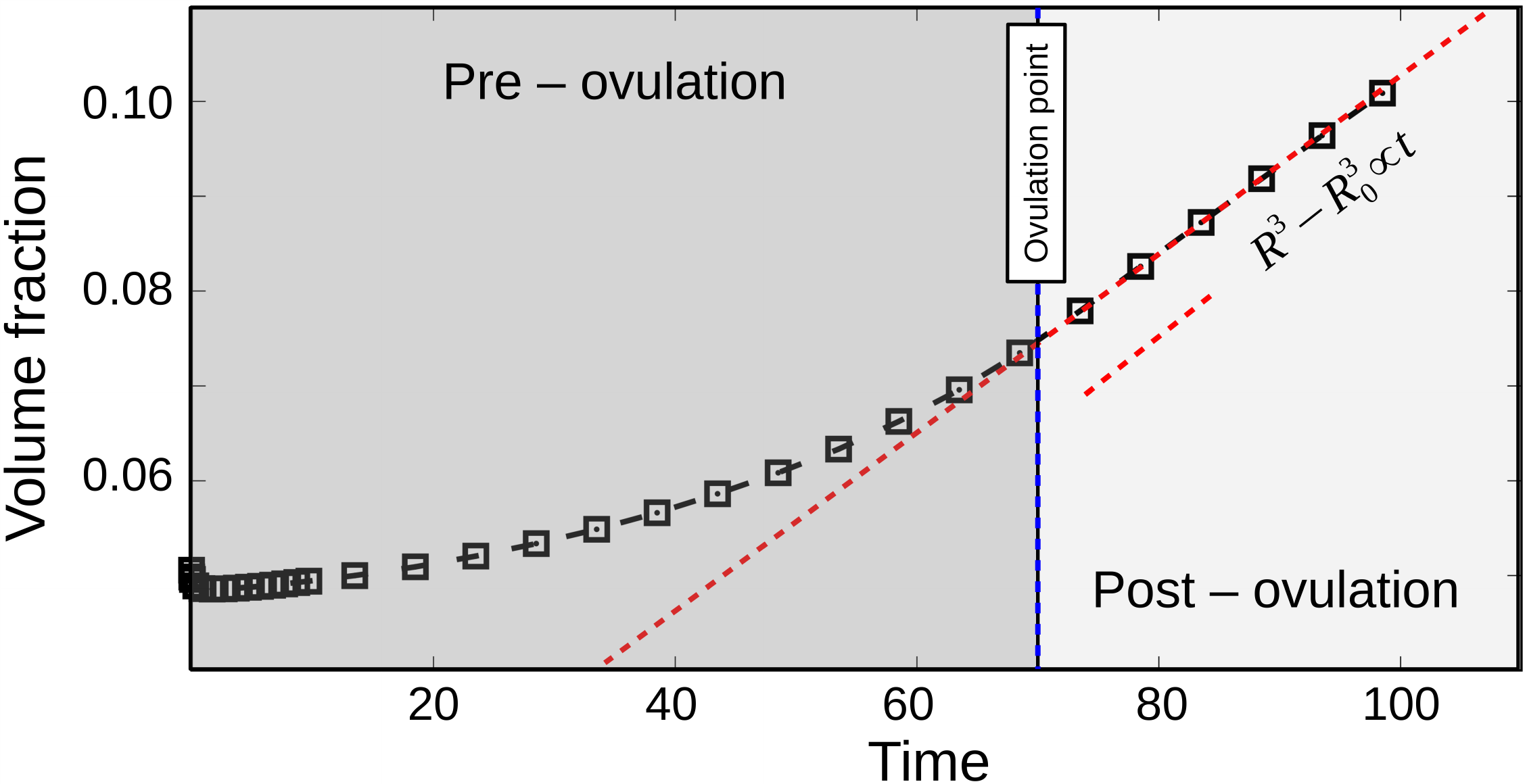}
    \end{tabular}
    \caption{ The increase in the volume of cementite at the longitudinal ends with time during the entire transformation of the rod of aspect ratio 20.
    \label{fig:fig9}}
\end{figure}

A characteristic feature of the transformation analysed in the present work is that the volume fractions of the phases remain unaltered throughout the evolution.
However, as elucidated earlier, considerable mass transfer is induced by the curvature-difference and other equilibration condition.
These activities ensuing the temporal evolution of the cementite rod are predominantly associated with the longitudinal ends which are fused with the grain boundaries.
Therefore, to understand the transformation rate, the progressive change in the volume fraction of cementite in the domain-section illustrated in Fig.~\ref{fig:fig7}a is ascertained and plotted in Fig.~\ref{fig:fig9}.
Since the different stages of the evolution are distinctly defined in the transformation of the rod of aspect ratio 20, this setup is chosen to analyse the kinetics.

Consistent with the temporal evolution of the shape elucidated in previous section, in the early stages, the change in the volume fraction of the precipitate at the termination is gradual.
The low rate of initial transformation is due to the uniform mass transfer from the remnant structure which establishes the characteristic termination morphology that spans across the grains.
Once the characteristic longitudinal ends begin to grow, a considerable difference in the curvature is introduced.
With the introduction of the curvature-difference, the termination volume increases at a faster rate, as shown in Fig.~\ref{fig:fig9}, when compared to the initial stages of the transformation.

The continued mass-transfer from a specific source in the remnant section of the rod, ultimately, results in ovulation.
The volume change in the termination structure is significantly more pronounced after the fragmentation, as shown in Fig.~\ref{fig:fig9}.
Since the transformation following the pinch-off is governed by the Ostwald ripening, the shape of the grain-boundary precipitate is assumed to be spherical and the radius $R$ is correspondingly determined from its volume.
With $R_o$ and $R$ representing the radius immediately following the ovulation and at a given instant, the relation $R^3-R_o^3$ is calculated and included in Fig.~\ref{fig:fig9}.
The proportional increase in $R^3-R_o^3$ with time indicates that the coarsening of the termination structure adheres to the power law.
Interestingly, the analytical relation pertaining to coarsening is recovered in spite of the characteristic morphology adopted by the grain-boundary cementite.

\subsection{Ovulation criterion and time}

\begin{figure}
    \centering
      \begin{tabular}{@{}c@{}}
      \includegraphics[width=0.75\textwidth]{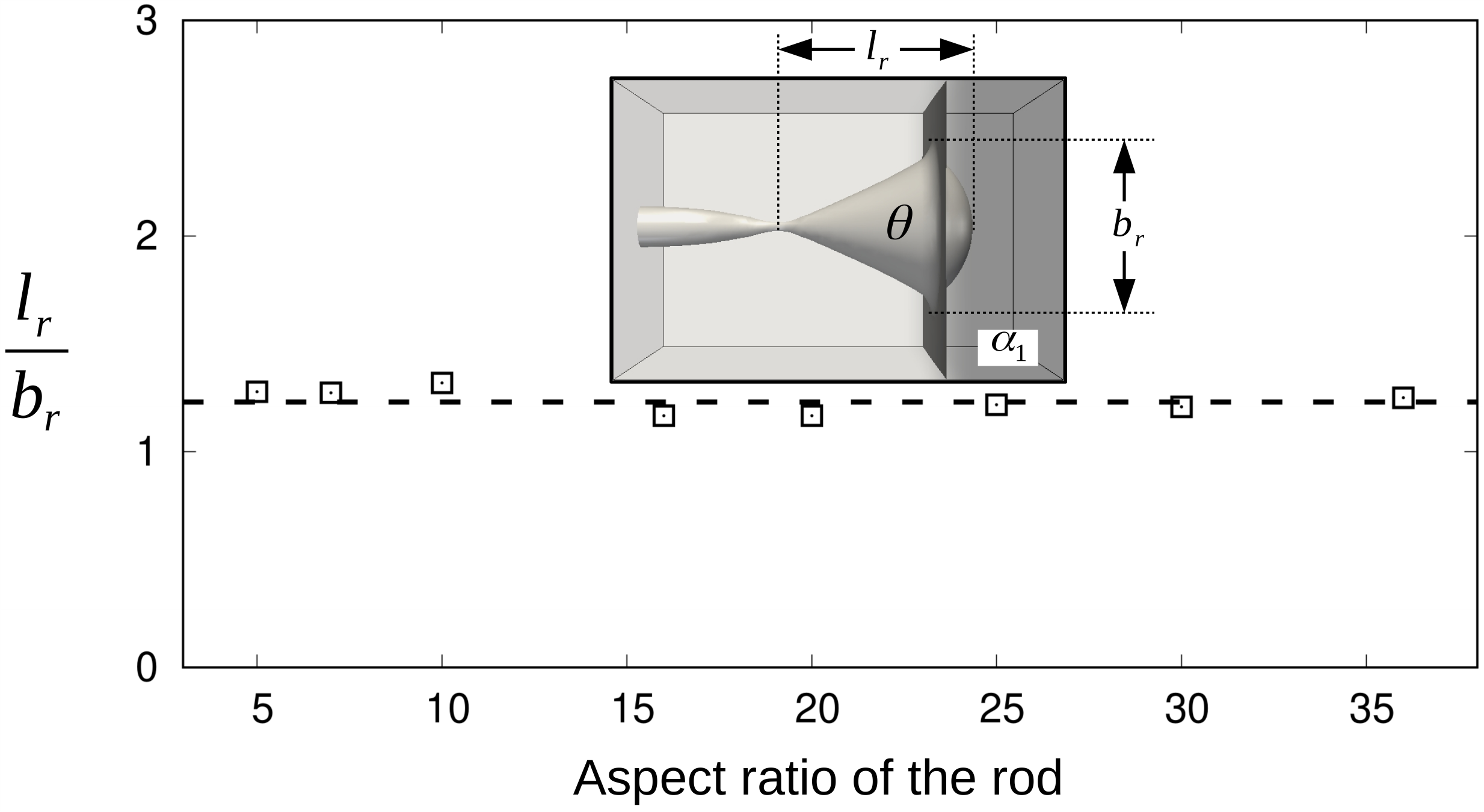}
    \end{tabular}
    \caption{ The ratio of the length ($l_r$) and width ($b_r$) of the termination structure assumed by the precipitate of different aspect ratio at the point of ovulation.
    \label{fig:fig6}}
\end{figure}

In Figs.~\ref{fig:fig4} and ~\ref{fig:fig5}, it is noticeable that with increase in the aspect ratio of the precipitate rod, the size of the satellite particle resulting from the ovulation proportionately increases.
In other words, the pinch-off site shifts proportionately from the centre with the increase in the size of the cementite rod.
Similar behaviour is exhibited by the isolated rods during spheroidization~\cite{amos2018phase2}. 
Moreover, in the isolated structures, it is identified that the shift in the ovulation site corresponds to a definite criterion which can be described based on the dimension of the longitudinal ridges at the point of pinch-off.
To unravel such criterion, which undergirds the fragmentation during the morphological transformation of rods in a polycrystalline setup, shape-instability of wide range structures with varying aspect ratio is examined.
The aspect ratio of the characteristic shape at the longitudinal ends is calculated as shown in the subset of Fig.~\ref{fig:fig6} and included in the same illustration. 

Fig.~\ref{fig:fig6} shows that, irrespective of the initial size of the rod, the ratio of the length and width of the characteristic termination is equal at the point of ovulation.
This equivalence in the termination aspect-ratio ($l_r/b_r$) during fragmentation is responsible for the proportional shift in the ovulation-site away from the centre with increase in the size of the rod.
This shift appropriately increases the size of the satellite structures with the initial aspect ratio of the rod.
Therefore, analogous to the evolution of the isolated structures, Fig.~\ref{fig:fig6} indicates that the shape-instability, particularly the fragmentation, of the rods in a multiphase setup is dictated by the ovulation criterion defined by the dimensions of the characteristic morphology at the termination.

\begin{figure}
    \centering
      \begin{tabular}{@{}c@{}}
      \includegraphics[width=0.75\textwidth]{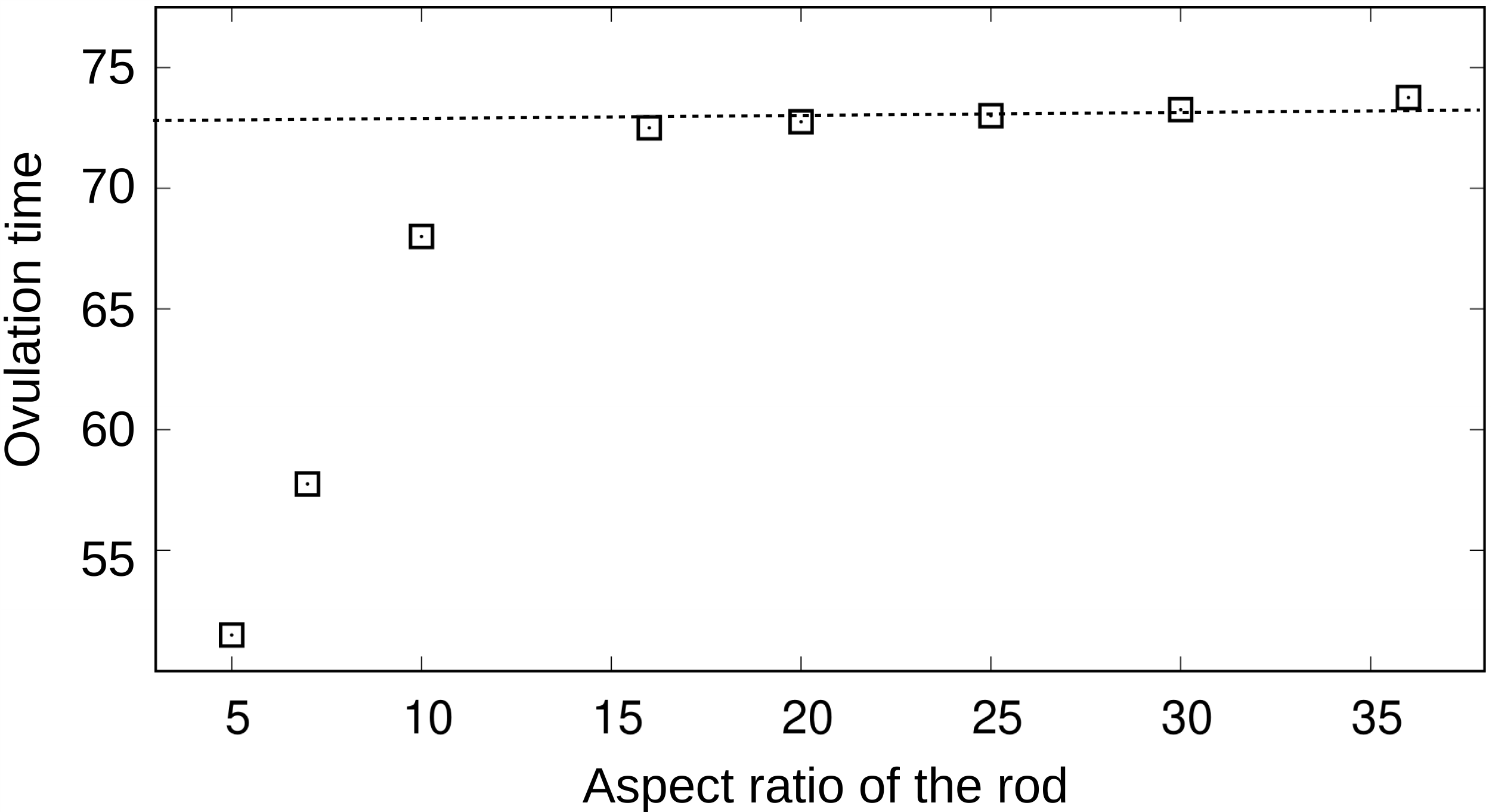}
    \end{tabular}
    \caption{ The time taken for the fragmentation of the continuous rods of different initial sizes during its evolution in a polycrystalline setup.
    \label{fig:fig8}}
\end{figure}

In addition to the ovulation criterion, another characteristic feature observed during the spheroidization of finite isolated rods pertains to the ovulation time.
It is noticed that, in both surface and volume diffusion governed transformation, the time taken for the initial pinch-off is seemingly equal in larger rods~\cite{nichols1976spheroidization,amos2018phase2}. 
However, in smaller structures the ovulation time varies.
The influence of initial size on the ovulation time of the rod in a polycrystalline setup is ascertained by monitoring the evolution.
The time for the fragmentation of the cementite structure of different aspect ratios is plotted in Fig.~\ref{fig:fig8}.
Akin to the evolution of the isolated finite-rods, the ovulation time increases with the initial size in the polycrystalline rods of aspect ratio upto 15.
However, in larger rods of aspect ratio 15 and above, the time taken for the pinch-off is apparently equal.

\section{Conclusion}

Morphological stability of the phases in a microstructure is analysed by considering a structure corroborating the shape and allowing it to evolve under an appropriate thermodynamic condition.
Generally, the system accommodating the structure comprises of the precipitate engulfed in the matrix.
Therefore, the driving force is predominantly governed by the inherent curvature-difference in the shape and the precipitate evolves to reduce the overall interfacial energy of the system.
Since isolated structures are observed in a microstructure and the influence of the neighbours considerably decrease with the distance, these investigations render critical insights on understanding the morphological stability of the phases.
However, microstructures of applicable materials are largely polycrystalline in nature, and grain boundaries which contribute to the nucleation during phase transformation, also play a vital role in the shape transformation.
Moreover, it is identified that, during pearlite spheroidization, the grain boundary assists in the formation of the isolated finite-cementite structures which later exhibits morphological changes governed by the inherent difference in the curvature.
To delineate the role of grain boundaries in the temporal evolution of three-dimensional rods, under chemical equilibrium, the stability of the precipitate in a representative polycrystalline setup is analysed.

The present phase-field analysis of the shape-instability of rods in a multiphase system unravels that the curvature-difference is not the sole governing factor.
In fact, the shape-changes at the early stages of the transformation are dictated by the energy-minimising events at the triple junctions which include the interaction between the phase and grain boundaries.
This initial morphological evolution establishes a characteristic morphology at the longitudinal ends of the rods, which eventually introduces curvature-difference.
The resulting mass transfer from the remnant section of the rod to the termination leads to pinch-off which fragments the continuous structure.
Depending on the initial size of the rod, the ovulation either occurs at the centre of the rod or at the foot of the morphologically-transformed terminations.
The latter, which is observed in the larger rods, yields the evolution of a satellite particle whose size varies with the initial aspect-ratio of the rod.
Although the fragmentation in the larger structures is invariably followed by coarsening, the isolated finite satellite structure evolves governed by the inherent difference in the curvature.
Therefore, the current work elucidates the sequence of events resulting in the formation of the isolate structures, which are predominantly observed during the morphological evolution of phases in a polycrystalline microstructure.
This grain-boundary assisted fragmentation deepens and adds to the current understanding of the shape-instability.

Since this investigation is primarily directed towards expounding the difference introduced by the polycrystalline setup on the energy-minimising shape-change exhibited by the rod, the entire analysis has been confined to a definite set of material parameters in a representative domain.
Exhaustive study analysing the morphological transformation under varied thermodynamical conditions in physical microstructural setups will be reported in the future.
For instance, anisotropy in the interfacial energy separating grains and phases would noticeably alter the geometrical configuration of the triple junction.
Consequently, the morphology of the cementite particles at the grain boundaries change. 
Understanding the effect of different interfacial energies on the kinetics and mechanism of the morphological evolution form a pivotal aspect of the forthcoming investigations.
Moreover, by extending the present approach through the incorporation of mechanical driving-forces, influences of elastic and plastic strains on the transformation will be examined

Despite the ability of the phase-field model to accommodate multicomponent system, the present work considers a binary Fe-C system.
Therefore, attempts are made to understand the shape-change in systems comprising of components with significantly different diffusion-coefficients, and with considerable disparity in the surface and volume diffusion.
The role of interdiffusivity in the kinetics of the transformation will also be addressed in the upcoming works.

\appendix
\numberwithin{equation}{section}
\makeatletter 
% "activate" the preparatory code, but for section-level headers only
\newcommand{\section@cntformat}{A \thesection:\ }
\makeatother

\section*{Appendix: Interface properties}

In this section, the interface properties rendered by the present formulation are discussed.
In a two-phase system, the overall energy-density under the current considerations is expressed as
\begin{align}\label{eq:2func}
\calF(\vphi,\n \vphi,\vv{c})=\epsilon \gamma |\n \phi(\vv{x},t)|^2 + \frac{\gamma}{\epsilon}\frac{16}{\pi^2}\phi(\vv{x},t)(1-\phi(\vv{x},t)) + f_{\text{chem}}(\vv{c}(\vv{x},t),\vphi).
\end{align}
Under the conditions of chemical equilibrium and a single spatial dimension, both, the contribution of the bulk phases as well as the influence of curvature, become insignificant.
Correspondingly, the system remains in a static condition with no temporal evolution of the phase-field.
The absence of phase-field evolution, 
\begin{align}\label{eq:1DphiEv}
\epsilon\tau\frac{\partial \phi(x,t)}{\partial t}=-\frac{\delta \mathcal{F}(\phi,\n_{x} \phi,\vv{c})}{\delta \phi}\equiv 0,
\end{align}
renders the relation
\begin{align}\label{eq:2DphiEv}
2\gamma\epsilon\frac{\partial^2 \phi}{\partial x^2}=\frac{\gamma}{\epsilon}\frac{16}{\phi^2}(1-2\phi).
\end{align}  
Both sides of Eqn.~\eqref{eq:2DphiEv} are integrated by including $\frac{\diff \phi}{\diff x}$, 
\begin{align}\label{eq:3DphiEv}
\int  \left( \frac{\diff \phi}{\diff x} \right ) \frac{\partial^2 \phi}{\partial x^2} \diff x= \int \frac{1}{\epsilon^2}\frac{16}{\pi^2}\frac{1-2\phi}{2} \left( \frac{\diff \phi}{\diff x}\right) \diff x.
\end{align}  
Upon integrating, above Eqn.\eqref{eq:3DphiEv} transforms to 
\begin{align}\label{eq:4DphiEv}
\left( \frac{\diff \phi}{\diff x} \right )^2= \frac{1}{\epsilon^2}\frac{16}{\pi^2} \phi(1-\phi).
\end{align}
The interface width $\Lambda$ is estimated by
\begin{align}\label{eq:intw1}
\Lambda=\int_{0}^{\Lambda}\diff x
\end{align}
Based on Eqn.~\eqref{eq:4DphiEv}, the width of the diffuse interface can be determined by
\begin{align}\label{eq:intw2}
\int_{0}^{\Lambda}\diff x = \epsilon\int_{0}^{1}\left[ \frac{16}{\pi^2} \phi(1-\phi) \right]^{-\frac{1}{2}} \diff \phi
\end{align}
Above Eqn.~\eqref{eq:intw2}, after integrating is written as
\begin{align}\label{eq:intw3}
\Lambda=\epsilon\frac{\pi}{4}\left [  \sin^{-1} (2\phi - 1) \right  ]_{0}^{1}.
\end{align}
From the above expression, for a given length parameter $\epsilon$, the width of the diffuse interface is
\begin{align}\label{eq:intw4}
\Lambda=\epsilon\frac{\pi^2}{4}.
\end{align}
The transition of the phase-field across the diffuse interface depends on the formulation of the interface contribution.
The interface profile can be ascertained by considering Eqn.~\eqref{eq:intw3}.
Accordingly, the respective relation is written as
\begin{align}\label{eq:intpro1_1}
\int_{0}^{x}\diff x=\epsilon\frac{\pi}{4}\sin{-1}(2\phi(x)-1).
\end{align} 
By re-arranging the terms, the interface profile for the present formulation is expressed as
\begin{align}\label{eq:intpro1_2}
\phi(x) = \frac{1}{2}+\frac{1}{2}\sin\frac{4}{\epsilon \pi}x.
\end{align}

\section*{Acknowledgements}
PGK Amos thanks the financial support of the German Research Foundation (DFG) under the project AN 1245/1.
The authors gratefully thank the financial support of the Helmholtz association through the programme HGF 34 ``EMR- Energy efficiency, materials and resources''. 
Pieces of this work were performed on the computational resource ForHLR II, funded by the Ministry of Science, Research and Arts of Baden-Wuerttemberg and the DFG.

\section*{Declaration of interest}
The authors have no competing interests.

\section*{References}

\bibliographystyle{elsarticle-num}
\bibliography{library.bib}
\end{document}